\documentclass[acmsmall]{acmart}

\usepackage{booktabs} 

\usepackage{cleveref}
\usepackage{epsfig}
\usepackage{float}
\usepackage{amsmath,amssymb,amsfonts}
\usepackage{algorithm}
\usepackage[noend]{algpseudocode}
\usepackage{bigstrut,multirow}
\usepackage{threeparttable}
\usepackage{subfigure}
\usepackage{url}

\acmJournal{TRETS}
\acmVolume{9}
\acmNumber{4}
\acmArticle{11}
\acmYear{2017}
\acmMonth{12}
\acmArticleSeq{11}

\setcopyright{usgovmixed}

\acmDOI{0000001.0000001}

\begin{document}
\title{[DL] A Survey of FPGA-Based Neural Network Inference Accelerator} 

\author{Kaiyuan Guo, Shulin Zeng, Jincheng Yu, Yu Wang {\scshape and} Huazhong Yang}
\affiliation{%
  \institution{Tsinghua University}
  \streetaddress{30 Shuangqing Rd, Haidian District}
  \city{Beijing}
  \state{Beijing}
  \postcode{100084}
  \country{China}
}
\email{gky15@mails.tsinghua.edu.cn, zengsl18@mails.tsinghua.edu.cn, yjc16@mails.tsinghua.edu.cn, yu-wang@mail.tsinghua.edu.cn, yanghz@tsinghua.edu.cn}

%
%

\begin{abstract}

Recent researches on neural network have shown significant advantage in machine learning over traditional algorithms based on handcrafted features and models. Neural network is now widely adopted in regions like image, speech and video recognition. But the high computation and storage complexity of neural network inference poses great difficulty on its application. CPU platforms are hard to offer enough computation capacity. GPU platforms are the first choice for neural network process because of its high computation capacity and easy to use development frameworks. 

On the other hand, FPGA-based neural network inference accelerator is becoming a research topic. With specifically designed hardware, FPGA is the next possible solution to surpass GPU in speed and energy efficiency. Various FPGA-based accelerator designs have been proposed with software and hardware optimization techniques to achieve high speed and energy efficiency. In this paper, we give an overview of previous work on neural network inference accelerators based on FPGA and summarize the main techniques used. An investigation from software to hardware, from circuit level to system level is carried out to complete analysis of FPGA-based neural network inference accelerator design and serves as a guide to future work. 

\end{abstract}

%
%



\maketitle

\renewcommand{\shortauthors}{K. Guo et al.}
\newcommand{\rev}[1]{{\color[rgb]{0,0,0}{#1}}}

\section{Introduction}\label{sec:introduction}

Recent research on Neural Network (NN) is showing great improvement over traditional algorithms in machine learning. Various network models, like convolutional neural network (CNN), recurrent neural network (RNN), have been proposed for image, video, and speech process. CNN~\cite{krizhevsky2012imagenet} improves the top-5 image classification accuracy on ImageNet~\cite{ILSVRC15} dataset from 73.8\% to 84.7\% in 2012 and further helps improve object detection~\cite{girshick2014rich} with its outstanding ability in feature extraction. RNN~\cite{hannun2014deep} achieves state-of-the-art word error rate on speech recognition. In general, NN features a high fitting ability to a wide range of pattern recognition problems. This ability makes NN a promising candidate for many artificial intelligence applications.

But the computation and storage complexity of NN models are high. In Table~\ref{tab:cnn_list}, we list the number of operations, number of parameters (add or multiplication), and top-1 accuracy on ImageNet dataset~\cite{ILSVRC15} of state-of-the-art CNN models. Take CNN as an example. The largest CNN model for a $224\times224$ image classification requires up to 39 billion floating point operations (FLOP) and more than 500MB model parameters~\cite{simonyan2014very}. As the computation complexity is proportional to the input image size, processing images with higher resolutions may need more than 100 billion operations. Latest work like MobileNet~\cite{Howard2017MobileNets} and ShuffleNet~\cite{zhang2017shufflenet} are trying to reduce the network size with advanced network structures, but with obvious accuracy loss. The balance between the size of NN models and accuracy is still an open question today. In some cases, the large model size hinders the application of NN, especially in power limited or latency critical scenarios.

\begin{table}[t]
    \centering
    \caption{Performance and resource utilization of state-of-the-art neural network accelerator designs}
    \begin{tabular}{c|c|c|c|c|c}
        \toprule
            & AlexNet\cite{krizhevsky2012imagenet} & VGG19\cite{simonyan2014very} & ResNet152\cite{he2016deep} & MobileNet\cite{Howard2017MobileNets} & ShuffleNet\cite{zhang2017shufflenet}\\\hline
        Year            & 2012      & 2014      & 2016      & 2017      & 2017      \\
        \# Param        & 60M       & 144M      & 57M       & 4.2M      & 2.36M     \\ 
        \# Operation    & 1.4G      & 39G       & 22.6G     & 1.1G      & 0.27G     \\ 
        Top-1 Accuracy  & 61.0\%    & 74.5\%    & 79.3\%    & 70.6\%    & 67.6\%    \\        
        \bottomrule
    \end{tabular}%
    \label{tab:cnn_list}%
  \end{table}%

Therefore, choosing a proper computation platform for neural-network-based applications is essential. A typical CPU can perform 10-100G FLOP per second, and the power efficiency is usually below 1GOP/J. So CPUs are hard to meet the high performance requirements in cloud applications nor the low power requirements in mobile applications. In contrast, GPUs offer up to 10TOP/s peak performance and are good choices for high performance neural network applications. Development frameworks like Caffe~\cite{jia2014caffe} and Tensorflow~\cite{abadi2016tensorflow} also offer easy-to-use interfaces which makes GPU the first choice of neural network acceleration. 

Besides CPUs and GPUs, FPGAs are becoming a platform candidate to achieve energy efficient neural network processing. With a neural network oriented hardware design, FPGAs can implement high parallelism and make use of the properties of neural network computation to remove additional logic. Algorithm researches also show that an NN model can be simplified in a hardware-friendly way while not hurting the model accuracy. Therefore FPGAs are possible to achieve higher energy efficiency compared with CPU and GPU. 

FPGA-based accelerator designs are faced with two challenges in performance and flexibility:
\begin{itemize}
    \item Current FPGAs usually support working frequency at 100-300MHz, which is much less than CPU and GPU. The FPGA's logic overhead for reconfigurability also reduces the overall system performance. A straightforward design on FPGA is hard to achieve high performance and high energy efficiency.
    \item Implementation of neural networks on FPGAs is much harder than that on CPUs or GPUs. Development framework like Caffe and Tensorflow for CPU and GPU is absent for FPGA.
\end{itemize}
 
Many designs addressing the above two problems have been carried out to implement energy efficient and flexible FPGA-based neural network accelerators. In this paper, we summarize the techniques proposed in these work from the following aspects:
\begin{itemize}
    \item We first give a simple model of FPGA-based neural network accelerator performance to analyze the methodology in energy efficient design.
    \item We investigate current technologies for high performance and energy efficient neural network accelerator designs. We introduce the techniques in both software and hardware level and estimate the effect of these techniques.
    \item We compare state-of-the-art neural network accelerator designs to evaluate the techniques introduced and estimate the achievable performance of FPGA-based accelerator design, which is at least $10\times$ better energy efficient than current GPUs.
    \item We investigate state-of-the-art automatic design methods of FPGA-based neural network accelerators. 
\end{itemize}

The rest part of this paper is organized as follows: Section~\ref{sec:preliminary} introduces the basic operations of neural networks and the background of FPGA-based NN accelerator. In section~\ref{sec:design_method}, we analyze the design target of NN accelerators and corresponding methods. Section~\ref{sec:software} and section~\ref{sec:hardware} review the techniques in NN model compression and accelerator design respectively. Section~\ref{sec:evaluation} compares existing designs and evaluate the techniques. Section~\ref{sec:flexibility} introduces the methods for a flexible accelerator design. Section~\ref{sec:conclusion} concludes this paper.

\section{Preliminary}\label{sec:preliminary}

Before discussing the system design for neural network acceleration, we first introduce the basic concepts of neural networks and the typical structure of FPGA-based NN accelerator design.

\subsection{Neural Network}

\begin{figure}[ht]
    \centering
    \includegraphics[width=1.0\columnwidth]{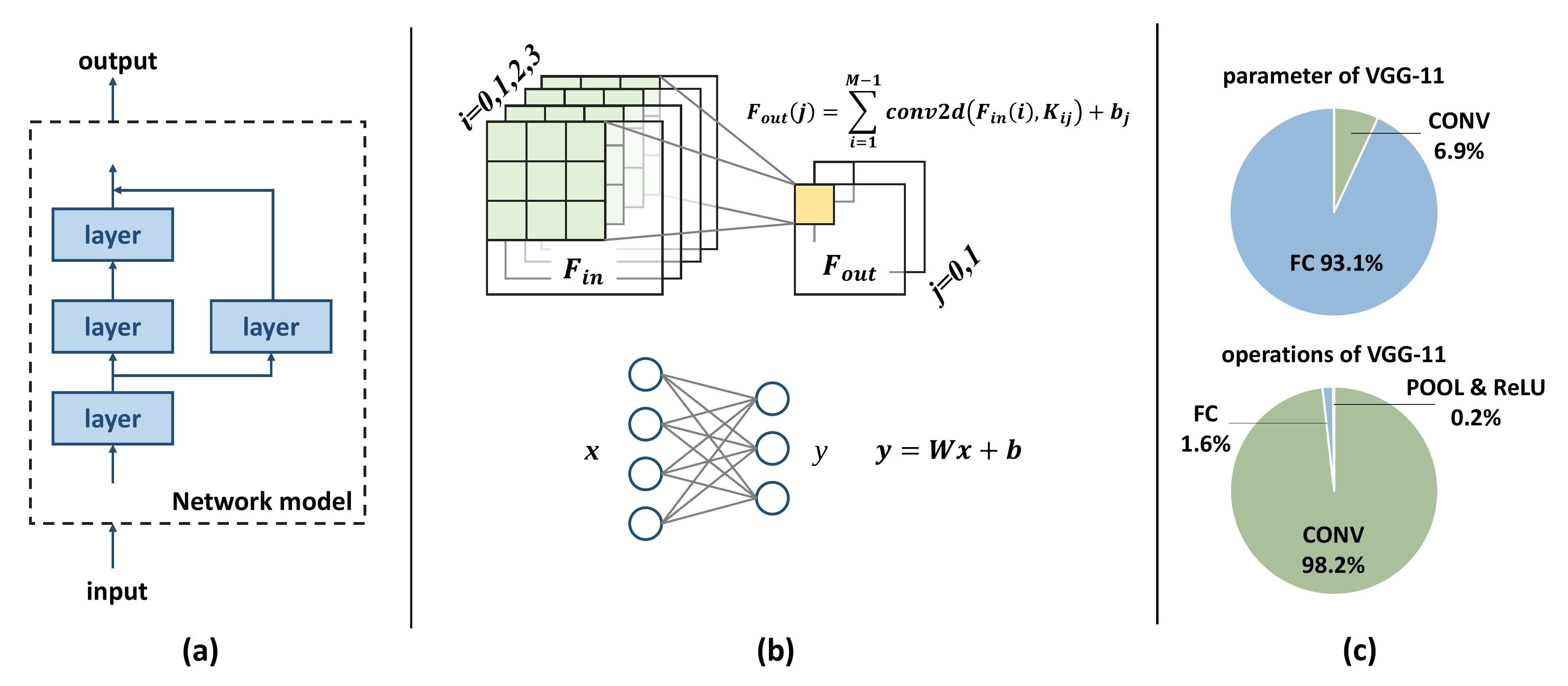}
    \caption{(a) Computation graph of a neural network model. (b) CONV and FC layers in NN model. (c) CONV and FC layers dominate the computation and parameter of a typical NN model: VGG11.}
    \label{fig:cnn_preliminary}
\end{figure}

In this section, we introduce the basic functions in a neural network. In this paper, we only focus on the inference of NN, which means using a trained model to predict or classify new data. The training process of NN is not discussed in this paper. A neural network model can be expressed as a directed graph shown in Figure~\ref{fig:cnn_preliminary}(a). Each vertex of the graph denotes a layer which conducts operations on data from a previous layer or input and generates results to the next layer or output. We refer the parameter of each layer as weights and the input/output of each layer as activations through this paper. 

Convolution (CONV) layers and fully connected (FC) layers are two common types of layers in NN models. The functions of these two layers are shown in Figure~\ref{fig:cnn_preliminary}(b). CONV layers conduct 2D convolutions on a set of input feature maps $F_{in}$ and add the results to get output feature maps $F_{out}$. FC layers receive a feature vector as input and conduct matrix-vector multiplications.

Besides CONV and FC layers, NN layers also have pooling, ReLU~\cite{krizhevsky2012imagenet}, concat~\cite{szegedy2015going}, element-wise~\cite{he2016deep} and other types of layers. But these layers contributes little to the computation and storage requirement of a neural network model. Figure~\ref{fig:cnn_preliminary}(c) shows the distribution of weights and operations in the VGG-11 model~\cite{simonyan2014very}. In this model, CONV and FC layers together contribute more than 99\% of the network's weights and operations, which is similar to most of the CNN models. Compared with CNN, RNN models~\cite{hannun2014deep, amodei2016deep} usually have no CONV layers and only FC layers contributes to most of the computation and storage. So most of the neural network acceleration systems focus on these two types of layers.

\subsection{FPGA-based Accelerator}\label{sec:preliminary:fpga}

In recent years, FPGA is becoming a promising solution for algorithm acceleration. Compared with CPU, GPU, and DSP platforms, for which the software and hardware are designed independently, FPGA enables the developers to implement only the necessary logic in hardware according to the target algorithm. By eliminating the redundancy in general hardware platforms, FPGAs can achieve higher efficiency. Application specific integrated circuits (ASICs) based solutions achieve even higher efficiency but requires much longer development cycle and higher cost. 

\begin{figure}[ht]
    \centering
    \includegraphics[width=0.9\columnwidth]{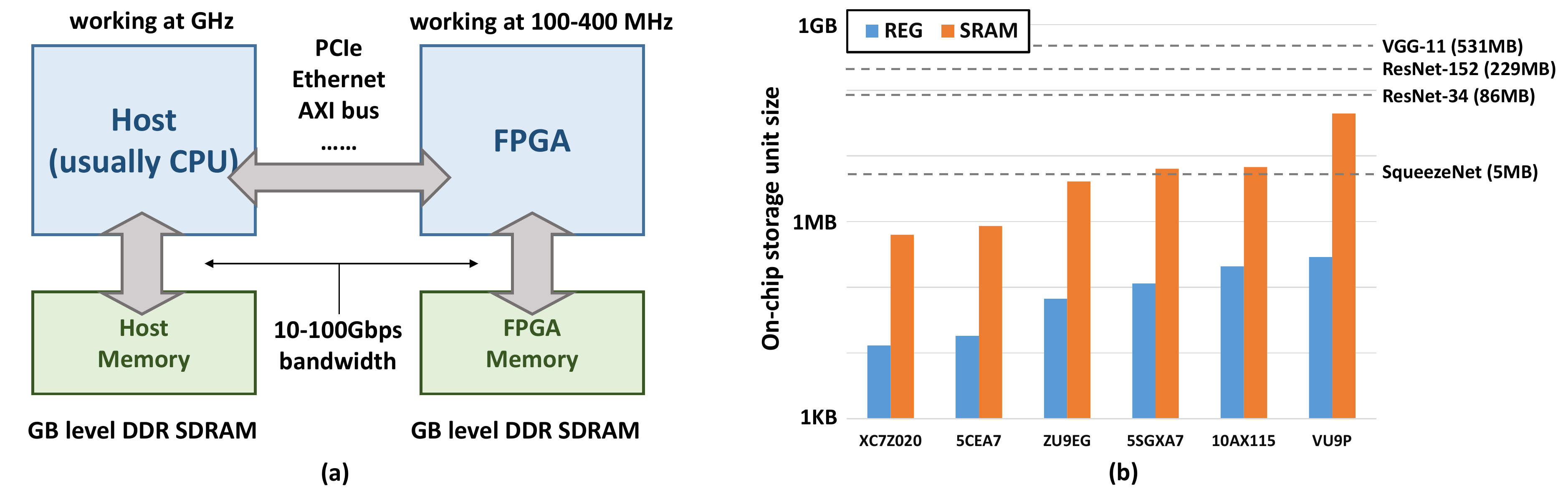}
    \caption{\rev{(a) A typical structure of an FPGA-based NN accelerator. (b) Gap between NN model size and the storage unit size on FPGAs. The bar chart compares the register and SRAM sizes on FPGA chips in different scales. The dotted line denotes the parameter sizes of different NN models with 32-bit floating point parameters.}}
    \label{fig:fpga_preliminary}
\end{figure}

For FPGA-based neural network accelerator, a typical architecture of the system is shown in Figure~\ref{fig:fpga_preliminary}(a). The system usually consists of a CPU host and an FPGA part. A pure FPGA chip usually works with a host PC/server through PCIe connections. SoC platforms (like the Xilinx Zynq Series) and Intel HARPv2~\cite{gupta2016accelerating} platform integrate the host and the FPGA in the same chip or package. Both the host and the FPGA can work with their own external memory and access each others' memory through the connection. Most of the designs implement NN accelerator on the FPGA part and control the 
accelerator with the software on the host.

Typical FPGA chips consist large on-chip storage units like registers and \rev{SRAM(Static Random-Access Memory)}, but still too small compared with NN models as shown in Figure~\ref{fig:fpga_preliminary}(b). Common models implement 100-1000MB parameters while the largest available FPGA chip implements <50MB on-chip SRAM. This gap requires that external memory like DDR SDRAM is needed. The bandwidth and power consumption of DDR limits the system performance.

The computation capacity of FPGA is relatively higher. Common FPGAs implement hundreds to thousands of DSP units, each of which can compute $18\times 27$ or $18\times 19$, achieving up to 10TFLOP/s (floating point operations per second) on the largest FPGAs. But for low-end FPGAs like Xilinx XC7Z020, this number is reduced to 20GFLOP/s, which is hard to support real-time video processing for applications on mobile platforms. 

Even faced with the above challenges, researchers have proposed a series of optimization methods from algorithm to architecture to design high performance NN accelerators on FPGA, which will be discussed in the following sections of this paper.

\section{Design Methodology and Criteria}\label{sec:design_method}

Before going into the details of the techniques used for neural network accelerators, we first give an overview of the design methodology. In general, the design target of a neural network inference accelerator includes the following two aspects: high speed (high throughput and low latency), and high energy efficiency. The symbols used in this section are listed in Table~\ref{tab:symbol}.

\begin{table}[htbp]
    \centering
    \begin{threeparttable}
        \caption{List of Symbols}\label{tab:symbol}%
        \begin{tabular}{l|p{0.6\columnwidth}|l} \toprule
        Symbol & Description & Unit \\ \hline
        $IPS$   & Throughput of the system, measured by the number of inference processed each second & $s^{-1}$ \\ \hline
        $W$     & Workload for each inference, measured by the number of operations$^*$ in the network, mainly addition and multiplication for neural network. & GOP \\ \hline
        $OPS_{peak}$ & Peak performance of the accelerator, measured by the maximum number of operations can be processed each second. & GOP/s \\ \hline
        $OPS_{act}$ & Run-time performance of the accelerator, measured by the number of operations processed each second. & GOP/s \\ \hline
        $\eta$   & Utilization ratio of the computation units, measured by the average ratio of working computation units in all the computation units during each inference. & - \\ \hline
        $f$ & Working frequency of the computation units. & GHz \\ \hline
        $P$ & Number of computation units in the hardware design. & - \\ \hline
        $L$ & Latency for processing each inference & s \\ \hline
        $C$ & Concurrency of the accelerator, measured by the number of inference processed in parallel & - \\ \hline 
        $Eff$   & Energy efficiency of the system, measured by the number of operations can be processed within unit energy. & GOP/J \\ \hline
        $E_{total}$ & Total system energy cost for each inference. & J \\ \hline 
        $E_{static}$ & Static energy cost of the system for each inference. & J \\ \hline
        $E_{op}$ & Average energy cost for each operation in each inference. & J \\ \hline
        $N_{x\_acc}$ & Number of bytes accessed from memory ($x$ can be SRAM or DRAM). & byte \\ \hline
        $E_{x\_acc}$ & Energy for accessing each byte from memory($x$ can be SRAM or DRAM). & J/byte \\ 
        \bottomrule
        \end{tabular}%
        \begin{tablenotes}
            \item[*] Each addition or multiplication is counted as 1 operation.
        \end{tablenotes}
    \end{threeparttable}
\end{table}%

\textbf{Speed}. The throughput of an NN accelerator can be expressed by equation~\ref{eqt:throughput}. The on-chip resource for a certain FPGA chip is limited. We can increase the peak performance by: 1. increasing the number of computation units $P$ by reducing the size of each computation unit and 2. increasing the working frequency $f$. Reducing the size of computation units can be achieved by sacrificing the data precision, which may hurt the model accuracy and requires hardware-software co-design. On the other hand, increasing working frequency is pure hardware design work. Corresponding techniques on software models and hardware are introduced in section~\ref{sec:software} and \ref{sec:hardware} respectively. A high utilization ratio $\eta$ is ensured by reasonable parallelism implementation and efficient memory system. The property of the target model, i.e. the data access pattern or data-computation ratio also affect if the hardware can be fully utilized at run-time. But most of the previous work targeting higher utilization ratio focus on the hardware side.

\begin{equation}\label{eqt:throughput}
    IPS = \frac{OPS_{act}}{W} = \frac{OPS_{peak} \times \eta}{W} = \frac{fP\times\eta}{W}
\end{equation}

Most of the FPGA-based NN accelerators compute different inputs one by one. Some designs process different inputs in parallel. So the latency of the accelerator is expressed as equation~\ref{eqt:latency}. Common concurrent design includes layer pipeline and batch processing. This is usually considered together with loop unrolling and will be introduced in section~\ref{sec:hardware:lu}. In this paper, we focus more on optimizing the throughput. As different accelerators may be evaluated on different NN models, a common criterion of speed is the $OPS_{act}$, which eliminates the effect of different network models to some extent.

\begin{equation}\label{eqt:latency}
    L = \frac{C}{IPS}
\end{equation}

\textbf{Energy Efficiency}. Energy efficiency ($Eff$) is another critical criteria to computing systems. For neural network inference accelerators, energy efficiency is defined as equation~\ref{eqt:efficiency}. Like throughput, we count the number of operations rather than the number of inference to eliminates the difference of workload $W$. If the workload for the target network is fixed, increasing the energy efficiency of a neural network accelerator means to reduce the total energy cost, $E_{total}$ to process each input. 

\begin{equation}\label{eqt:efficiency}
    Eff = \frac{W}{E_{total}}
\end{equation}
    
\begin{equation}\label{eqt:energy}
    E_{total} \approx W\times E_{op} + N_{SRAM\_acc}\times E_{SRAM\_acc} + N_{DRAM\_acc}\times E_{DRAM\_acc} + E_{static}
\end{equation}

The total energy cost mainly comes from 2 parts: computation and memory access, which is expressed in equation~\ref{eqt:energy}. The first item in equation~\ref{eqt:energy} is the dynamic energy cost for computation. Given a certain network, the workload $W$ is fixed. Researchers have been focusing on optimizing the NN models by quantization (narrowing the bit-width used for computation) to reduce $E_{op}$ or sparsification (setting more weights to zeros) to skip the multiplications with these zeros to reduce $N_{op}$, which follows similar rules as for throughput optimization. 

The second and third item in equation~\ref{eqt:energy} is the dynamic energy cost for memory access. As shown in section~\ref{sec:preliminary:fpga}, FPGA-based NN accelerator usually works with an external DRAM. We separate the memory access energy into DRAM part and SRAM part. $N_{x\_acc}$ can be reduced by quantization, sparsification, efficient on-chip memory system, and scheduling method. Thus these methods help reduce dynamic memory energy. Corresponding methods will be introduced in section~\ref{sec:hardware:sys}. The unit energy $E_{x\_acc}$ can hardly be reduced given a certain FPGA platform.

The fourth item $E_{static}$ denotes the static energy cost of the system. This energy cost can hardly be improved given the FPGA chip and the scale of the design.

From the analysis of speed and energy, we see that neural network accelerator involves both optimizations on NN models and hardware. In the following sections, we will introduce previous work in these two aspects respectively.

\section{Hardware Oriented Model Compression}\label{sec:software}

As introduced in section~\ref{sec:design_method}, the design of energy efficient and fast neural network accelerator can benefit from the optimization of NN models. A larger NN model usually results in higher model accuracy. This means it is possible to trade the model accuracy for the hardware speed or energy cost. Neural network researchers are designing more efficient network models from AlexNet~\cite{krizhevsky2012imagenet} to ResNet~\cite{he2016deep}, SqueezeNet~\cite{iandola2016squeezenet} and MobileNet~\cite{Howard2017MobileNets}. Latest work tries to directly optimize the processing latency by searching a good network structure~\cite{tan2018mnasnet} or skip some layers at run-time to save computation~\cite{wang2017skipnet}. Within these methods, the main differences between the handcrafted/generated networks are the size of and the connections between each layer. The basic operations are the same and the differences hardly affect the hardware design. For this reason, we will not focus on these techniques in this paper. But designers should consider using these techniques to optimize the target network.

Other methods try to achieve the tradeoff by compressing existing NN models. They try to reduce the number of weights or reduce the number of bits used for each activation or weight, which help lower down the computation and storage complexity. Corresponding hardware designs can benefit from these NN model compression methods. In this section, we investigate these hardware oriented network model compression methods.

\subsection{Data Quantization}\label{sec:software:quant}
One of the most commonly used methods for model compression is the quantization of the weights and activations. The activations and weights of a neural network are usually represented by floating point data in common developing frameworks. Recent work tries to replace this representation with low-bit fixed-point data or even a small set of trained values. On the one hand, using fewer bits for each activation or weight helps reduce the bandwidth and storage requirement of the neural network processing system. On the other hand, using a simplified representation reduce the hardware cost for each operation. The benefit of hardware will be discussed in detail in section~\ref{sec:hardware}. Two kinds of quantization methods are discussed in this section: linear quantization and non-linear quantization.

\subsubsection{Linear Quantization}
Linear quantization finds the nearest fixed-point representation of each weight and activation. The problem with this method is that the dynamic range of floating-point data greatly exceeds that for fixed-point data. Most of the weights and activations will suffer from overflow or underflow. Qiu et al.~\cite{qiu2016going} finds that the dynamic range of the weights and activations in a single layer is much more limited and differs across different layers. Therefore they assign different fractional bit-widths to the weights and activations in different layers. To decide the fractional bit-width of a set of data, i.e. the activations or weights of a layer, the data distribution is first analyzed. A set of possible fractional bit-widths are chosen as candidate solutions. Then the solution with the best model performance on training data set is chosen. In~\cite{qiu2016going}, the optimized solution of a network is chosen layer by layer to avoid an exponential design space exploration. Wang et al.~\cite{wang2018design} try to use large bit-width for only the first and last layer and quantize the middle layers to ternary or binary. The method needs to increase the network size to keep high accuracy but still brings hardware performance improvement.  Guo et al.~\cite{guo2017angel} choose to fine-tune the model after the fractional bit-width of all the layers are fixed.  

The method of choosing a fractional bit-width equals to scale the data with a scaling factor of $2^k$. Li et al.~\cite{li2016ternary} scales the weights with trained parameter $W^l$ for each layer and quantize the weights with 2-bit data, representing $W^l$, 0 and $-W^l$. The activations in this work are not quantized. So the network still implements 32-bit floating point operations. Zhou et al.~\cite{zhou2016dorefa} further quantize the weights of a layer with only 1 bit to $\pm s$, where $s=E(|w^l|)$ is the expectation of the absolute value of the weights of this layer. Linear quantization is also applied to the activations in this work.

\subsubsection{Non-linear Quantization}
Compared with linear quantization, non-linear quantization independently assigns values to different binary codes. The translation from a non-linear quantized code to its corresponding value is thus a look-up table. This kind of methods helps further reduce the bit-width used for each activation or weight. Chen et al.~\cite{chen2015compressing} assign each of the weight to an item in the look-up table by a pre-defined hash function and train the values in look-up tables. Han et al.~\cite{han2015deep} assign the values in look-up tables to the weights by clustering the weights of a trained model. Each look-up table value is set as the cluster centre and further fine-tuned with training data set. This method can compress the weights of state-of-the-art CNN models to 4-bit without accuracy loss. Zhu et al.~\cite{zhu2016trained} propose the ternary-quantized network where all the weights of a layer are quantized to three values: $W^n$, 0, and $W^p$. Both the quantized value and the correspondence between weights and look-up table are trained. This method sacrifices less than $2\%$ accuracy loss on ImageNet dataset on state-of-the-art network models. The weight bit-width is reduced from 32-bit to 2-bit, which means about $16\times$ model size compression.

\subsubsection{Comparison}
\begin{figure}[ht]
    \centering
    \includegraphics[width=1.0\columnwidth]{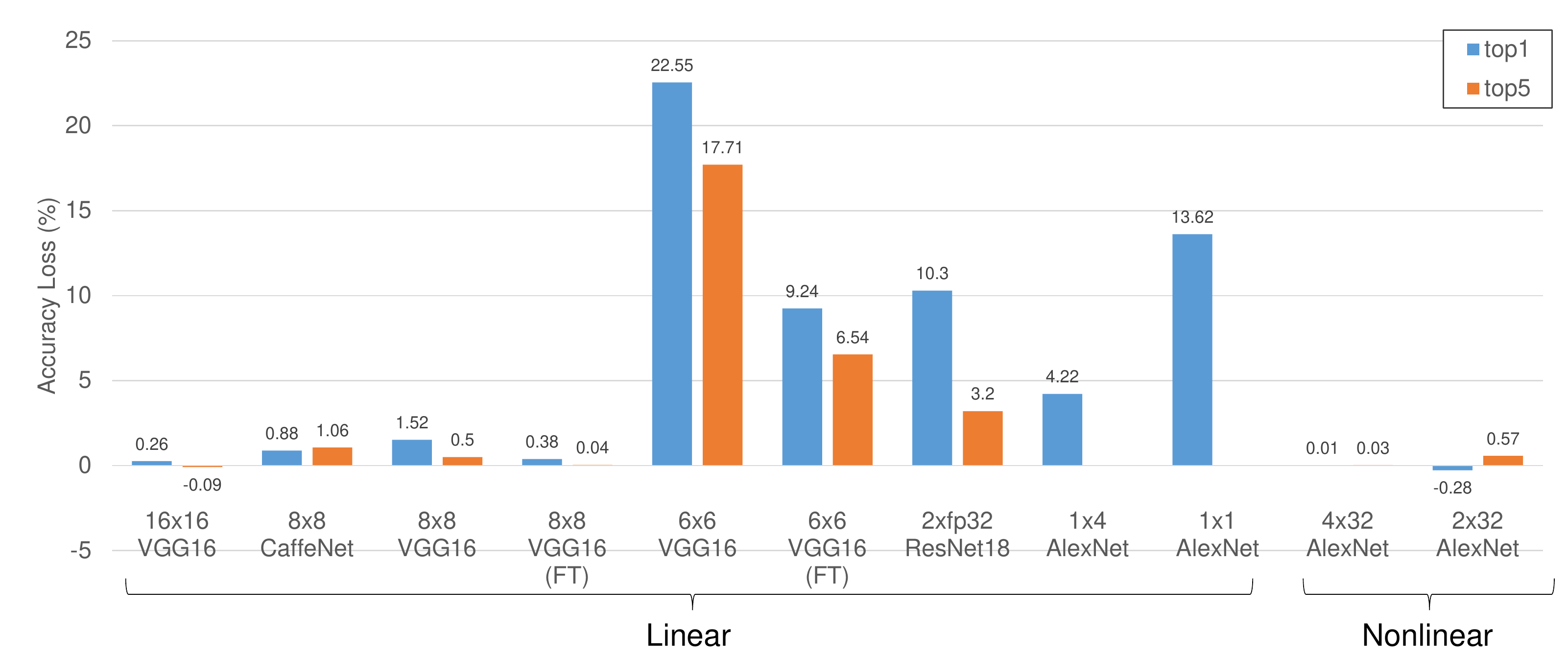}
    \caption{Comparison between different quantization methods from \cite{qiu2016going, guo2017angel, han2015deep, zhu2016trained, zhou2016dorefa, li2016ternary}. The quantization configuration is expressed as (weight bit-width)$\times$(activation bit-width). The "(FT)" denotes that the network is fine-tuned after a linear quantization.}
    \label{fig:quantization}
\end{figure}

We compare some typical quantization methods from~\cite{qiu2016going, guo2017angel, han2015deep, zhu2016trained, zhou2016dorefa, li2016ternary} in Figure~\ref{fig:quantization}. All the quantization results are tested on ImageNet data set and the absolute accuracy loss compared with corresponding baseline floating point models is recorded. Comparing different methods on different models is a little bit unfair. But it still gives some insights. For linear quantization, 8-bit is a clear bound to ensure negligible accuracy loss. With 6 or fewer bits, using fine-tune or even training each weight from the beginning will cause noticeable accuracy degradation. If we require that $1\%$ accuracy loss is within the acceptable range, linear quantization with at least $8\times 8$ configuration and the listed non-linear quantization are available. We will further discuss the performance gain of quantization in section~\ref{sec:hardware}.

\subsection{Weight Reduction}\label{sec:software:wr}
Besides narrowing the bit-width of activations and weights, another method for model compression is to reduce the number of weights. One kind of method is to approximate the weight matrix with a low-rank representation. Qiu et al.~\cite{qiu2016going} compress the weight matrix $W$ of an FC layer with singular value decomposition. An $m\times n$ weight matrix $W$ is replaced by the multiplication of two matrices $A_{m\times p}B_{p\times n}$. For a sufficiently small $p$, the total number of weights is reduced. This work compresses the largest FC layer of VGG network to $36\%$ of its original size with $0.04\%$ classification accuracy degradation. Zhang et al.~\cite{zhang2015efficient} use a similar method for convolution layers and takes the effect of the following non-linear layer into the decomposition optimization process. The proposed method achieves $4\times$ speed up on state-of-the-art CNN model targeting at ImageNet, with only $0.9\%$ accuracy loss.

Pruning is another kind of method to reduce the number of weights. This kind of methods directly remove the zeros in weights or remove those with small absolute values. The challenge in pruning is the tradeoff between the ratio of zero weights and the model accuracy. One solution is the application of lasso method, which applies L1 normalization to the weights during training. Liu et al.~\cite{liu2015sparse} apply the sparse group-lasso method on the AlexNet~\cite{krizhevsky2012imagenet} model. $90\%$ weights are removed after training with less than $1\%$ accuracy loss. Another solution is to prune the zero weights during training. Han et al.~\cite{han2015deep} directly remove the weights of a network which are zero or have small absolute value. The left weights are then fine-tuned with the training dataset to recover accuracy. Experimental results on AlexNet show that $89\%$ weights can be removed while keeping the model accuracy.

The hardware gain from weight reduction is the reciprocal of the compression ratio. According to the above results, the potential speed improvement from weight reduction is up to $10\times$.

\section{Hardware Design: Efficient Architecture}\label{sec:hardware}

In this section, we investigate the hardware level techniques used in state-of-the-art FPGA-based neural network accelerator design to achieve high performance and high energy efficiency. We classify the techniques into three levels: computation unit level, loop unrolling level, and system level.

\subsection{Computation Unit Designs}\label{sec:hardware:cu}

Computation unit level design affects the peak performance of the neural network accelerator. The available resource of an FPGA chip is limited. A smaller computation unit design means more computation units and higher peak performance. A carefully designed computation unit array can also increase the working frequency of the system and thus improve peak performance.

\subsubsection{Low Bit-width Computation Unit}\label{sec:hardware:cu:lbu}
Reduce the number of bit-width for computation is a direct way to reduce the size of computation units. The feasibility of using fewer bits comes from the quantization methods as introduced in section~\ref{sec:software:quant}. Most of the state-of-the-art FPGA designs replace the 32-bit floating-point units with fixed-point units. Podili et al.~\cite{podili2017fast} implement 32-bit fixed-point units for the proposed system. 16-bit fixed-point units are widely adopted in \cite{qiu2016going, li2016high, xiao2017exploring, guan2017fp, zhang2016caffeine}. ESE~\cite{han2017ese} adopts 12-bit fixed-point weight and 16-bit fixed-point neurons design. Guo et al.~\cite{guo2017angel} use 8-bit units for their design on embedded FPGA. Recent work is also focusing on extremely narrow bit-width design. Prost-Boucle et al.~\cite{prost2017scalable} implements 2-bit multiplication with 1 LUT for ternary networks. Experiments in \cite{nurvitadhi2016accelerating} show that FPGA implementation of Binarized Neural Network (BNN) outperforms that on CPU and GPU. Though BNN suffers from accuracy loss, many designs explore the benefit of using 1-bit data for computation~\cite{li20177, nakahara2017batch, zhao2017accelerating, umuroglu2017finn, nakahara2017fully, jiao2017accelerating, moss2017high, yang2018fully, ghasemzadehrebnet}.

The designs mentioned above focus on computation units for linear quantization. For non-linear quantization, translating the data back to full precision for computation still costs many resources. Samragh et al.~\cite{samragh2017customizing} propose the factorized coefficients based dot product implementation. As the possible values of weights are quite limited for non-linear quantization, the proposed computation unit accumulates the multipliers for each possible weight value and calculate the result as the weighted sum of the values in look-up tables. In this way, the multiplication needed for one output neuron equals to the number of values in look-up table. The multiplications are replaced by random-addressed accumulations.

Most of the designs use one bit-width through the process of a neural network. Qiu et al.~\cite{qiu2016going} finds that neurons and weights in FC layers can use fewer bits compared with CONV layers while the accuracy is maintained. Heterogeneous computation units are used in the designs of \cite{zhao2017accelerating, guo2017bit}.

The size of computation units of different bit-widths is compared in Table~\ref{tab:mac}. Three kinds of implementations are tested: separate multiplier and adder with logic resource on Xilinx FPGA, multiply-add function with DSP units on Xilinx FPGA, and multiply-add function with DSP units on Altera FPGA. The resource consumption is the synthesis result by Vivado 2018.1 targeting Xilinx XCKU060 FPGA and Quartus Prime 16.0 targeting Altera Arria 10 GX1150 FPGA. The pure logic modules and the floating-point multiply and add modules are generated with IP core. The fixed-point multiply and add modules are implemented with $A*B+C$ in Verilog and automatically mapped to DSP by Vivado/Quartus.

We first give an overview of the size of the computation units by logic-only implementations. By compressing the weights and activations from 32-bit floating-point number to 8-bit fixed-point number, the multiplier and the adder are scaled down to about 1/10 and 1/50 respectively. Using 4-bit or smaller operators can bring further advantage but also incur significant accuracy loss as introduced in section~\ref{sec:software:quant}. 

Recent FPGAs consist of a large number of DSP units, each of which implements hard multiplier, pre-adder and accumulator core. The basic pattern of NN computation, multiplication and sum, also fits into this design. So we also test the multiply and add function implemented with DSP units. Because of the different DSP architectures, we test on both Xilinx and Altera platforms. Compared with the 32-bit floating-point function, fixed-point functions with narrow bit-width still shows an advantage in resource consumption. But for Altera FPGA, this advantage is not obvious because the DSP units natively support floating-point operations. 

Fixed-point functions with 16-or-less-bit fixed-point data are well fit into 1 DSP unit on either Xilinx or Altera FPGA. This shows that quantization hardly benefits the hardware if we use narrower bit-width like 8 or 4 in the aspect of computation. The problem is that the wide multipliers and adders in DSP units are underutilized in these cases. Nguyen et al.~\cite{nguyen2017double} propose the design to implement two narrow bit-width fixed-point multiplication with a single wide bit-width fixed-point multiplier. In this design, two multiplications, $AB$ and $AC$, are executed in the form of $A(B<<k+C)$. If $k$ is sufficiently large, the bits for $AB$ and $AC$ does not overlap in the multiplication result and can be directly separated. The design in~\cite{nguyen2017double} implements two 8-bit multiplications with one $25\times 18$ multiplier, where $k$ is 9. Similar methods can be applied to other bit-width and DSPs.


\begin{table}[htbp]
    \centering
    \caption{FPGA resource consumption comparison for multiplier and adder with different types of data.}
      \begin{tabular}{l|rr|rr|rrr|rr} \toprule
      \multirow{2}[4]{*}{} & \multicolumn{4}{c|}{Xilinx Logic} & \multicolumn{3}{c|}{Xilinx DSP} & \multicolumn{2}{c}{Altera DSP} \\ \cline{2-10}         
       & \multicolumn{2}{c|}{multiplier} & \multicolumn{2}{c|}{adder} & \multicolumn{3}{c|}{multiply \& add} & \multicolumn{2}{c}{multiply \& add} \\ \cline{2-10}
              & LUT   & FF    & LUT   & FF    & LUT   & FF    & DSP   & ALM   & DSP \\ \hline
      fp32    & 708   & 858   & 430   & 749   & 800   & 1284  & 2     & 1     & 1 \\ 
      fp16    & 221   & 303   & 211   & 337   & 451   & 686   & 1     & 213   & 1 \\ 
      fixed32 & 1112  & 1143  & 32    & 32    & 111   & 64    & 4     & 64    & 3 \\ 
      fixed16 & 289   & 301   & 16    & 16    & 0     & 0     & 1     & 0     & 1 \\ 
      fixed8  & 75    & 80    & 8     & 8     & 0     & 0     & 1     & 0     & 1 \\ 
      fixed4  & 17    & 20    & 4     & 4     & 0     & 0     & 1     & 0     & 1 \\ 
      \bottomrule
      \end{tabular}
    \label{tab:mac}
\end{table}

\subsubsection{Fast Convolution Method}\label{sec:hardware:cu:fcu}
For CONV layers, the convolution operations can be accelerated by alternative algorithms. Discrete Fourier Transformation (DFT) based fast convolution is widely adopted in digital signal processing. Zhang et al.~\cite{zhang2017frequency} propose a 2D DFT based hardware design for efficient CONV layer execution. For an $F\times F$ filter convolved with $K\times K$ filter, DFT converts the $(F-K+1)^2K^2$ multiplications in the space domain to $F^2$ complex multiplications in the frequency domain. For a CONV layer with $M$ input channel and $N$ output channel, $MN$ times of frequency domain multiplications and $(M+N)$ times DFT/IDFT are needed. The conversion of convolution kernels is once for all. So the domain conversion process is of low cost for CONV layers. This technique does not work for CONV layers with stride>1 or $1\times 1$ convolution. Ding et al.~\cite{ding2017c} suggest that a block-wise circular constraint can be applied to the weight matrix. In this way, the matrix-vector multiplication in FC layers are converted to a set of 1D convolutions and can be accelerated in the frequency domain. This method can also be applied to CONV layers by treating the $K\times K$ convolution kernels as $K\times K$ matrices and is not limited by $K$ or stride.

Frequency domain methods require complex number multiplication. Another kind of fast convolution involves only real number multiplication~\cite{winograd1980arithmetic}. The convolution of a 2D feature map $F_{in}$ with a kernel $K$ using Winograd algorithm is expressed by equation~\ref{eqt:winograd}.
\begin{equation}\label{eqt:winograd}
    F_{out} = A^T[(GF_{in}G^T)\odot(BF_{in}B^T)]A
\end{equation}
$G$, $B$ and $A$ are transformation matrix which only related to the sizes of kernel and feature map. $\odot$ denotes an element-wise multiplication of two matrices. For a $4\times 4$ feature map convolved with a $3\times 3$ kernel, the transformation matrices are described as follows:
\begin{equation*}
    G = \left[
        \begin{array}{ccc}
            1           & 0            & 0           \\
            \frac{1}{2} & \frac{1}{2}  & \frac{1}{2} \\
            \frac{1}{2} & -\frac{1}{2} & \frac{1}{2} \\
            0           & 0            & 1
        \end{array}    
    \right] \quad
    B = \left[
        \begin{array}{cccc}
            1 & 0  & -1 & 0 \\
            0 & 1  & 1  & 0 \\
            0 & -1 & 1  & 0 \\
            0 & 1  & 0  & -1
        \end{array}
    \right] \quad
    A = \left[
        \begin{array}{cc}
            1 & 0  \\
            1 & 1  \\
            1 & -1 \\
            0 & -1 
        \end{array}
    \right]
\end{equation*}
Multiplication with transformation matrices $A, B$ and $G$ induce only a small number of shift and addition because of the special matrix entries. In this case, the number of multiplication is reduced from 36 to 16.The most commonly used Winograd transformation is for $3\times 3$ convolutions in \cite{lu2017evaluating, xiao2017exploring}. 

The theoretical performance gain from fast convolution depends on the convolution size. Limited by the on-chip resource and the consideration of flexibility, current designs are not choosing large convolution sizes. Existing work point out that up to $4\times$ theoretical performance gain can be achieved by fast convolution with FFT~\cite{zhang2017frequency} or Winograd~\cite{lu2017evaluating} with reasonable kernel sizes. Zhuge et al.~\cite{zhuge2018face} even try to use both FFT and Winograd methods in their design to fit different kernel sizes in different layers.

\subsubsection{Frequency Optimization Methods}
All the above techniques introduced targets at increasing the number of computation units within a certain FPGA. Increasing the working frequency of the computation units also improves the peak performance.

Latest FPGAs support 700-900MHz DSP theoretical peak working frequency. But existing designs usually work at 100-400MHz~\cite{qiu2016going, guo2017angel, zhang2016caffeine, ma2017optimizing, zhang2017improving}. As claimed in \cite{wu2017high}, the working frequency is limited by the routing between on-chip SRAM and DSP units. The design in \cite{wu2017high} uses different working frequencies for DSP units and surrounding logic. Neighbor slices to each DSP unit are used as local RAMs to separate the clock domain. The prototype design in \cite{wu2017high} achieves the peak DSP working frequency at 741MHz and 891MHz on FPGA chips of different speed grades. \rev{Xilinx has also proposed the CHaiDNN-v2~\cite{chai_dnn} and xfDNN~\cite{xfdnn} with this technique and achieves up to 700MHz DSP working frequency. Compared with existing designs for which the frequency is within 300MHz, this technique brings at least $2\times$ peak performance gain.}

\subsection{Loop Unrolling Strategies}\label{sec:hardware:lu}
CONV layers and FC layers contribute to most of the computations and storage requirement of a neural network as introduced in section~\ref{sec:preliminary}. We express the CONV layer function in Figure~\ref{fig:cnn_preliminary}(b) as nested loops in Algorithm~\ref{alg:conv}. To make the code clear to read, we merge the loops along $x$ and $y$ directions for feature maps and 2-D convolution kernels respectively. An FC layer can be expressed as a CONV layer with feature map and kernel both of size $1\times 1$. Besides the loops in Algorithm~\ref{alg:conv}, we also call the parallelism of the process of multiple inputs as a batch. As we treat FC layers and CONV layers all as nested loops, the loop unrolling strategy can be applied both in CNN accelerators and RNN accelerators. But as the case for FC layers are rather simple, we tend to use CNN as examples in this section.

\begin{algorithm}  
    \caption{Convolution Layer}
    \label{alg:conv}
    \begin{algorithmic}[1]
        \Require feature map $F_{in}$ of size $M\times Y\times X$; 
                 convolution kernel $Ker$ of size $N\times M\times K\times K$;
                 bias vector $b$ of size $N$ 
        \Ensure  feature map $F_{out}$
        \Function {ConvLayer}{$F_{in}, Ker$}  
            \State Let $F_{out} \gets $ zero array of size $N\times(Y-K+1)\times(X-K+1)$  
            \For{$n=1$; $n<N$; $n++$} \Comment Output channel loop
                \For{$m=1$; $m<M$; $m++$} \Comment Input channel loop
                    \For{each $(y, x)$ within $(Y-K+1, X-K+1)$} \Comment Feature map loop
                        \For{each $(ky, kx)$ within $(K, K)$} \Comment Kernel loop
                            \State $F_{out}[n][y][x] += F_{in}[m][y-ky+1][x-kx+1] * K[n][m][ky][kx]$
                        \EndFor
                    \EndFor
                \EndFor
                \State $F_{out}[n] += b[n]$
            \EndFor
            \State \Return{$F_{out}$}
        \EndFunction  
        
    \end{algorithmic}  
\end{algorithm}

\subsubsection{Choosing Unroll Parameters}

To parallelize the execution of the loops, we unroll the loops and parallelize the process of a certain number of iterations on hardware. The number of the parallelized iterations on hardware is called the unroll parameter. Inappropriate unroll parameter selection may lead to serious hardware underutilization. Take a single loop as an example. Suppose the trip count of the loop is $M$ and the parallelism is $m$. The utilization ratio of the hardware is limited by $m/M\lceil M/m\rceil$. If $M$ is not divisible by $m$, then the utilization ratio is less than 1. For processing an NN layer, the total utilization ratio will be the product of the utilization ratio on each of the loops.



For a CNN model, the loop dimension varies greatly among different layers. For a typical network used on ImageNet classification like ResNet~\cite{he2016deep}, the channel numbers vary from 3 to 2048; the feature map sizes vary from $224\times 224$ to $7\times 7$, the convolution kernel sizes vary from $7\times 7$ to $1\times 1$. Besides the underutilization problem, loop unrolling also affect the datapath and on-chip memory design. Thus loop unrolling strategy is a key feature for a neural network accelerator design. 

Various work are proposed focusing on how to choose the unroll parameters. Zhang et al.~\cite{zhang2015optimizing} propose the idea of unrolling the input channel and output channel loops and choose the optimized unroll parameter by design space exploration. Along these two loops, there is no input data cross-dependency between neighboring iterations. So no multiplexer is needed to route data from the on-chip buffer to computation units. But the parallelism is limited as $7\times 64=448$ multipliers. For larger parallelism, this solution is easy to suffer from the underutilization problem. Ma et al.~\cite{ma2017optimizing} further extends the design space by allowing parallelism on the feature map loop. The parallelism reaches $1\times 16\times 14\times 14=3136$ multipliers. A shift register structure is used to route feature map pixels to the computation units.

The kernel loop is not chosen in the above work because kernel sizes vary greatly. Motamedi et al~\cite{motamedi2016design} use kernel unrolling on AlexNet. Even with $3\times 3$ unrolling for the $11\times 11$ and $5\times 5$ kernels, the overall system performance still reaches 97.4\% of its peak performance for the convolution layers. For certain networks like VGG~\cite{simonyan2014very}, only $3\times 3$ convolution kernels are used. Another reason to unroll kernel loop is to achieve acceleration with fast convolution algorithms. Design in \cite{zhang2017frequency} implements fully parallelized frequency domain multiplication on $4\times 4$ feature map and $3\times 3$ kernel. Lu et al.~\cite{lu2017evaluating} implement Winograd algorithm on FPGA with a dedicated pipeline for equation~\ref{eqt:winograd}. The convolution of a $6\times 6$ feature map with a $3\times 3$ kernel is fully parallelized.

The above solutions are only for a single layer. But there is hardly a one-size-fits-all solution for a whole network, especially when we need high parallelism. Designs in \cite{li2016high, liu2016automatic, zhang2018dnnbuilder} propose fully pipelined structures with each layer a pipe stage. As each layer is executed with an independent part of the hardware and each part is small, loop unrolling method can be easily chosen. This method is memory consuming because ping-pong buffers are needed between adjacent layers for the feature maps. Agressive design with binarized weights~\cite{yang2018fully} can fit into FPGA better. Design in \cite{zhang2016energy} is similar but implemented on FPGA clusters to resolve the scalability problem. Shen et al.~\cite{shen2016overcoming} and Lin et al.~\cite{lin2018lcp} group the layers of a CNN by the loops' trip count and map each group onto one hardware module. These solutions can be treated as unrolling the batch loop because different inputs are processed in parallel on different layer pipeline stages. The design in \cite{lu2017evaluating} implements parallelized batch both within a layer and among different layers. 

Most of the current designs follow one of the above methods for loop unrolling. A special kind of design is for sparse neural networks. Han et al.~\cite{han2017ese} propose the ESE architecture for sparse LSTM network acceleration. Unlike processing a dense network, all the computation units will not work synchronously. This causes difficulty in sharing data between different computation units. ESE implements only the output channel (the output neurons of the FC layers in LSTM) loop unrolling within a layer to simplify hardware design and parallelize batch process.

\subsubsection{Data Transfer and On-chip Memory Design}

Besides the high parallelism, the on-chip memory system should efficiently offer the necessary data to each computation units every cycle. To implement high parallelism, neural network accelerators usually reuse data among a large number of computation units. Simply broadcasting data to different computation units leads to large fan-out and high routing cost and thus reduce the working frequency. Wei et al.~\cite{wei2017automated} use the systolic array structure in their design. The shared data are transferred from one computation unit to the next in a chain mode. So the data is not broadcasted, and only local connections between different computation units are needed. The drawback is the increase in latency. The loop execution order is scheduled accordingly to cover the latency. Similar designs are adopted in~\cite{aydonat2017opencl, ma2017optimizing}. 

For software implementation on GPU, the im2col function is commonly used to map 2D convolution as a matrix-vector multiplication. This method incurs considerable data redundancy and can hardly be applied to the limited on-chip memory of FPGAs. Qiu et al.~\cite{qiu2016going} uses the line buffer design to achieve the $3\times 3$ sliding window function for 2-d convolution with only two lines of duplicated pixels. 

\subsection{System Design}\label{sec:hardware:sys}

\begin{figure}[t]
    \centering
    \includegraphics[width=0.8\columnwidth]{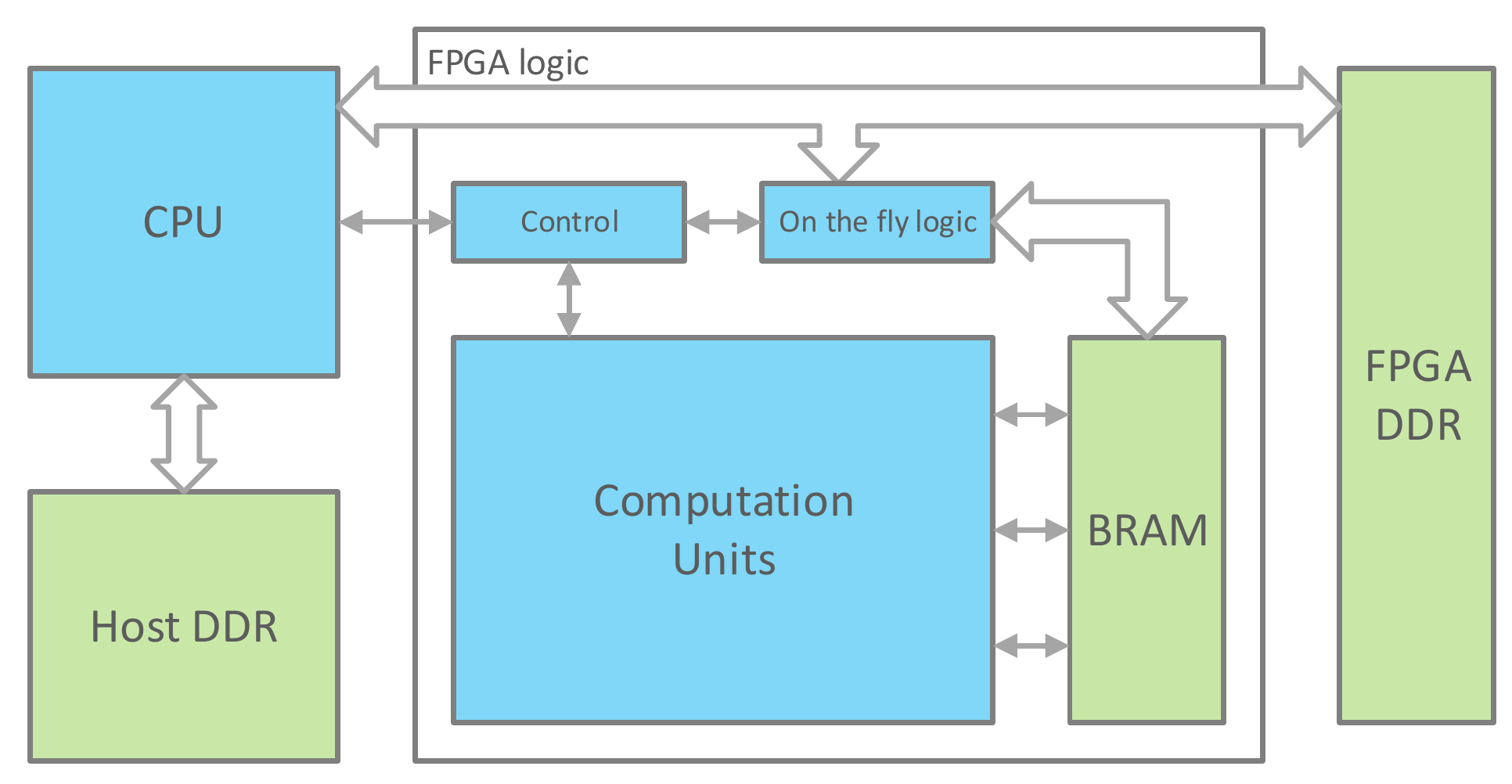}
    \caption{Block graph of a typical FPGA-based neural network accelerator system}
    \label{fig:sys}
\end{figure}

A typical FPGA-based neural network accelerator system is shown in Figure~\ref{fig:sys}. The logic part of the whole system is denoted by the blue boxes. The host CPU issues workload or commands to the FPGA logic part and monitors its working status. On the FPGA logic part, a controller is usually implemented to communicate with the host and generates control signals to all the other modules on FPGA. The controller can be an FSM or an instruction decoder. The on the fly logic part is implemented for certain designs if the data loaded from external memory needs preprocess. This module can be data arrangement module, data shifter~\cite{qiu2016going}, FFT module~\cite{zhang2017frequency}, etc. The computation units are as discussed in section~\ref{sec:hardware:cu} and section~\ref{sec:hardware:lu}. As introduced in section~\ref{sec:preliminary:fpga}, on-chip SRAM of an FPGA chip is too limited compared with the large NN models. So for common designs, a two-level memory hierarchy is used with DDR and on-chip memory. 

\subsubsection{\rev{Roofline Model}} 
\begin{figure}[h]
    \centering
    \includegraphics[width=0.6\columnwidth]{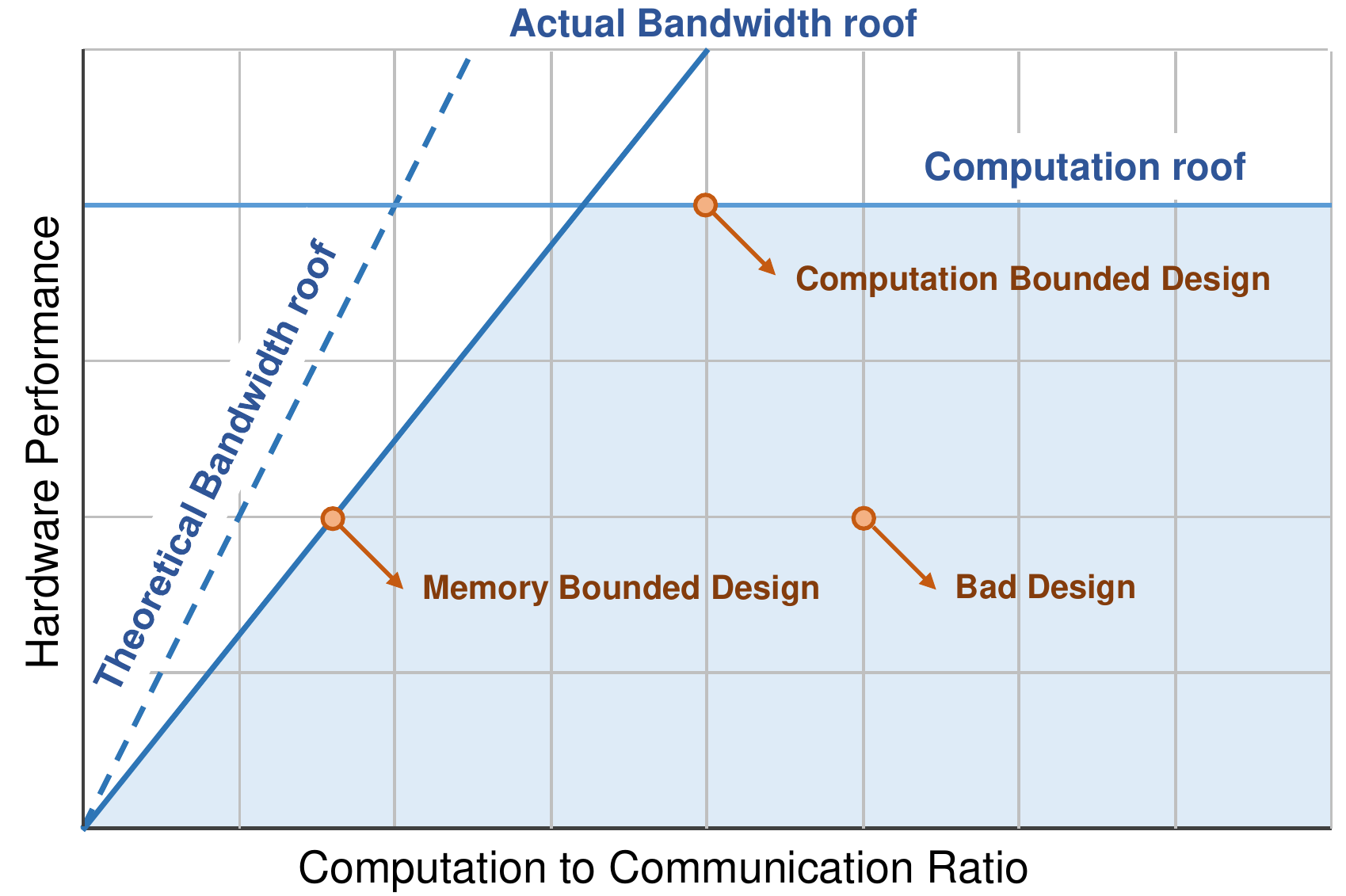}
    \caption{An example of the roofline model. The shaded part denotes the valid design space given bandwidth and resource limitation.}
    \label{fig:roofline}
\end{figure}

From the system level, the performance of a neural network accelerator is limited by two factors: the on-chip computation resource and the off-chip memory bandwidth. Various researches have been proposed to achieve the best performance within a certain off-chip memory bandwidth. Zhang et al.~\cite{zhang2015optimizing} introduce the roofline model in their work to analyze whether a design is memory bounded or computation bounded. An example of a roofline model is shown in Figure~\ref{fig:roofline}.

The figure uses the computation to communication (CTC) ratio as the $x$-axis and hardware performance as the $y$-axis. CTC is the number of operations that can be executed with a unit size of memory access. Each hardware design can be treated as a point in the figure. So $y/x$ equals to the bandwidth requirement of the design. The available bandwidth of a target platform is limited and can be described as the theoretical bandwidth roof in Figure~\ref{fig:roofline}. But the actual bandwidth roof is below the theoretical roof because the achievable bandwidth of DDR depends on the data access pattern. Sequential DDR access achieves much higher bandwidth than random access. The other roof is the computation roof, which is limited by the available resource on FPGA.

\subsubsection{\rev{Loop Tiling and Interchange}} A higher CTC ratio means the hardware is more likely to achieve the computation bound. Increasing the CTC ratio also reduce DDR access, which significantly saves energy according to~\cite{vlsi_energy}. In section~\ref{sec:hardware:lu}, we have discussed the loop unrolling strategies to increase the parallelism while reducing the waste of computation for a certain network. When the loop unrolling strategy is decided, the scheduling of the rest part of the loops decides how the hardware can reuse data with on-chip buffer. This involves loop tiling and loop interchange strategy.

Loop tiling is a higher level of loop unrolling. All the input data of a loop tile will be stored on-chip, and the loop unrolling hardware kernel works on these data. A larger loop tile size means that each tile will be loaded from external memory to on-chip memory fewer times. Loop interchange strategy decides the processing order of the loop tiles. External memory access happens when the hardware is moving from one tile to the next. Neighboring tile may share a part of data. For example in a CONV layer, neighboring tile can share input feature map or the weights. This is decided by the execution order of the loops. 

In ~\cite{zhang2015optimizing, ma2017optimizing}, design space exploration is done on all the possible loop tiling sizes and loop orders. Many designs also explore the design space with some of the loop unrolling, tiling and loop order is already decided~\cite{motamedi2016design, qiu2016going}. Shen et al.~\cite{shen2017escher} also discuss the effect of batch parallelism over the CTC for different layers. This is a loop dimension not focused on in previous work.

All the above work give one optimized loop unrolling strategy and loop order for a whole network. Guo et al.~\cite{guo2017angel} implements flexible unrolling and loop order configuration for different layers with an instruction interface. The data arrangement in on-chip buffers is controlled through instructions to fit with different feature map sizes. This means the hardware can always fully utilize the on-chip buffer to use the largest tiling size according to on-chip buffer size. This work also proposes the "back and forth" loop execution order to avoid total on-chip data refresh when an innermost loop finishes.

\subsubsection{\rev{Cross-Layer Scheduling}} Alwani et al.~\cite{alwani2016fused} address the external memory access problem by fusing two neighboring layers together to avoid the intermediate result transfer between the two layers. This strategy helps reduce 95\% off-chip data transfer with extra 20\% on-chip memory cost. Even software program gains $2\times$ speedup with this scheduling strategy. Yu et al.~\cite{Yu2017Instruction} realize this idea on a single-layer accelerator design by modifying the order of execution through an instruction interface.

\subsubsection{\rev{Regularize Data Access Pattern}} Besides increasing CTC, increasing the actual bandwidth roof helps improve the achievable performance with a certain CTC ratio. This is achieved by regularizing the DDR access pattern. The common feature map formats in the external memory include $NCHW$ or $CHWN$, where $N$ means the batch dimension, $C$ means the channel dimension, $H$ and $W$ means the feature map $y$ and $x$ dimension. Using any of these formats, a feature map tile may be cut into small data blocks stored in discontinuous addresses. Guan~\cite{guan2017fp} suggest that a channel-major storage format should be used for their design. This format avoids data duplication while long DDR access burst is ensured. Qiu et al.~\cite{qiu2016going} propose a feature map storage format that arranges the $H\times W$ feature map into $(HW/rc)$ tile blocks of size $r\times c$. So the write burst size can be increased from $c/2$ to $rc/2$.

\section{Evaluation}\label{sec:evaluation}

In this section, we compare the performance of state-of-the-art neural network accelerator designs and try to evaluate the techniques mentioned in section~\ref{sec:software} and section~\ref{sec:hardware}. We mainly reviewed the FPGA-based designs published in the top FPGA conferences (FPGA, FCCM, FPL, FPT), EDA conferences (DAC, ASPDAC, DATE, ICCAD), architecture conferences (MICRO, HPCA, ISCA, ASPLOS) since 2015. Because of the diversity in the adopted techniques, target FPGA chips, and experiments, we need a trade-off between the fairness of comparison and the number of designs we can use. In this paper, we pick the designs with 1) whole system implementation; 2) experiments on real NN models with reported speed, power, and energy efficiency.

The designs used for comparison are listed in Table~\ref{tab:hardware_list}. For data format, the "INT A/B" means that activations are A-bit fixed-point data and weights are B-bit fixed-point data. We also investigate the resource utilization and draw advice to both accelerator designers and FPGA manufacturers.

\begin{table}[htbp]
    \centering
    \caption{Performance and resource utilization of state-of-the-art neural network accelerator designs}
    \begin{tabular}{r|c|c|c|c|ccc|c}
        \toprule
        \multicolumn{1}{c|}{} & Data  & Speed & Power & Eff. & \multicolumn{3}{c|}{Resource(\%)} & \multirow{2}[4]{*}{FPGA chip} \\ 
        \multicolumn{1}{c|}{} & Format & (GOP/s) & (W)   & (GOP/J) & DSP   & logic & BRAM  &  \\
        \hline
            \cite{nakahara2017fully}    & 1bit      & 329.47    & 2.3   & 143.2 & 1     & 34    & 11    & Zynq XC7Z020 \\ 
            \cite{moss2017high}         & 1bit      & 40770     & 48    & 849.38 &   -   &   -   &   -   & GX1155 \\ 
            \cite{jiao2017accelerating} & 2bit      & 410.22    & 2.26  & 181.51 & 41   & 83    & 38    & Zynq XC7Z020 \\ 
            \cite{guo2017angel}         & INT8      & 84.3      & 3.5   & 24.1  & 87    & 84    & 89    & XC7Z020 \\ 
            \cite{suda2016throughput}   & INT16/8   & 117.8     & 19.1  & 6.2   & 13    & 22    & 65    & 5SGSD8 \\ 
            \cite{liu2016automatic}     & INT16/8   & 222.1     & 24.8  & 8.96  & 40    & 27    & 40    & XC7VX690T \\ 
            \cite{ma2017optimizing}     & INT16/8   & 645.25    & 21.2  & 30.43 & 100   & 38    & 70    & GX1150 \\             
            \cite{han2017ese}           & INT16/12  & 2520      & 41    & 61.5  & 54    & 89    & 88    & XCKU060 \\ 
            \cite{venieris2017fpgaconvnet} & INT16  & 12.73     & 1.75  & 7.27  & 95    & 67    & 6     & XC7Z020 \\ 
            \cite{qiu2016going}         & INT16     & 136.97    & 9.63  & 14.22 & 89    & 84    & 87    & XC7Z045 \\ 
            \cite{xiao2017exploring}    & INT16     & 229.5     & 9.4   & 24.42 & 92    & 71    & 83    & XC7Z045 \\ 
            \cite{zhang2016caffeine}    & INT16     & 354       & 26    & 13.6  & 78    & 81    & 42    & XC7VX690T \\ 
            \cite{guan2017fp}           & INT16     & 364.4     & 25    & 14.6  & 65    & 25    & 46    & 5SGSMD5 \\ 
            \cite{li2016high}           & INT16     & 565.94    & 30.2  & 22.15 & 60    & 63    & 65    & XC7VX690T \\ 
            \multirow{2}[2]{*}{\cite{Shen2018Towards}} & \multirow{2}[2]{*}{INT16} & 431  & 25 & 17.1  & 42 & 56 & 52    & XC7VX690T \\ 
             &  & 785 & 26 & 30.2 & 53 & 8.3  & 30 & XCVU440 \\ 
            \multirow{2}[2]{*}{\cite{zhang2016energy}} & \multirow{2}[0]{*}{INT16} & \multirow{2}[0]{*}{1280.3} & \multirow{2}[0]{*}{160} & \multirow{2}[0]{*}{8} & \multirow{2}[0]{*}{-} & \multirow{2}[0]{*}{-} & \multirow{2}[0]{*}{-} & XC7Z020+ \\
            &       &       &       &       &       &       &       & XC7VX690T$\times$6 \\ 
            \cite{zhang2017improving}   & INT16     & 1790      & 37.46 & 47.8  & 91    & 43    & 53    & GX1150 \\ 
            \cite{lu2017evaluating}     & INT16     & 2940.7    & 23.6  & 124.6 &   -   &   -   &   -   & ZCU102 \\ 
            \cite{aydonat2017opencl}    & FP16      & 1382      & 45    & 30.7  & 97    & 58    & 92    & GX1150 \\ 
            \cite{podili2017fast}       & INT32     & 229       & 8.04  & 28.5  & 100   & 84    & 18    & Stratix V \\ 
            \cite{guan2017fpga}         & FP32      & 7.26      & 19.63 & 0.37  & 42    & 65    & 52    & XC7VX485T \\ 
            \cite{zhang2015optimizing}  & FP32      & 61.62     & 18.61 & 3.3   & 80    & 61    & 50    & XC7VX485T \\ 
            \cite{zhang2017frequency}   & FP32      & 123.5     & 13.18 & 9.37  & 88    & 85    & 64    & Stratix V \\ 
            \cite{zhang2017improving}   & FP32      & 866       & 41.73 & 20.75 & 87    & -     & 46    & GX1150 \\ 
            \bottomrule
        \end{tabular}%
    \label{tab:hardware_list}%
  \end{table}%

\begin{figure}[ht]
    \centering
    \includegraphics[width=1.0\columnwidth]{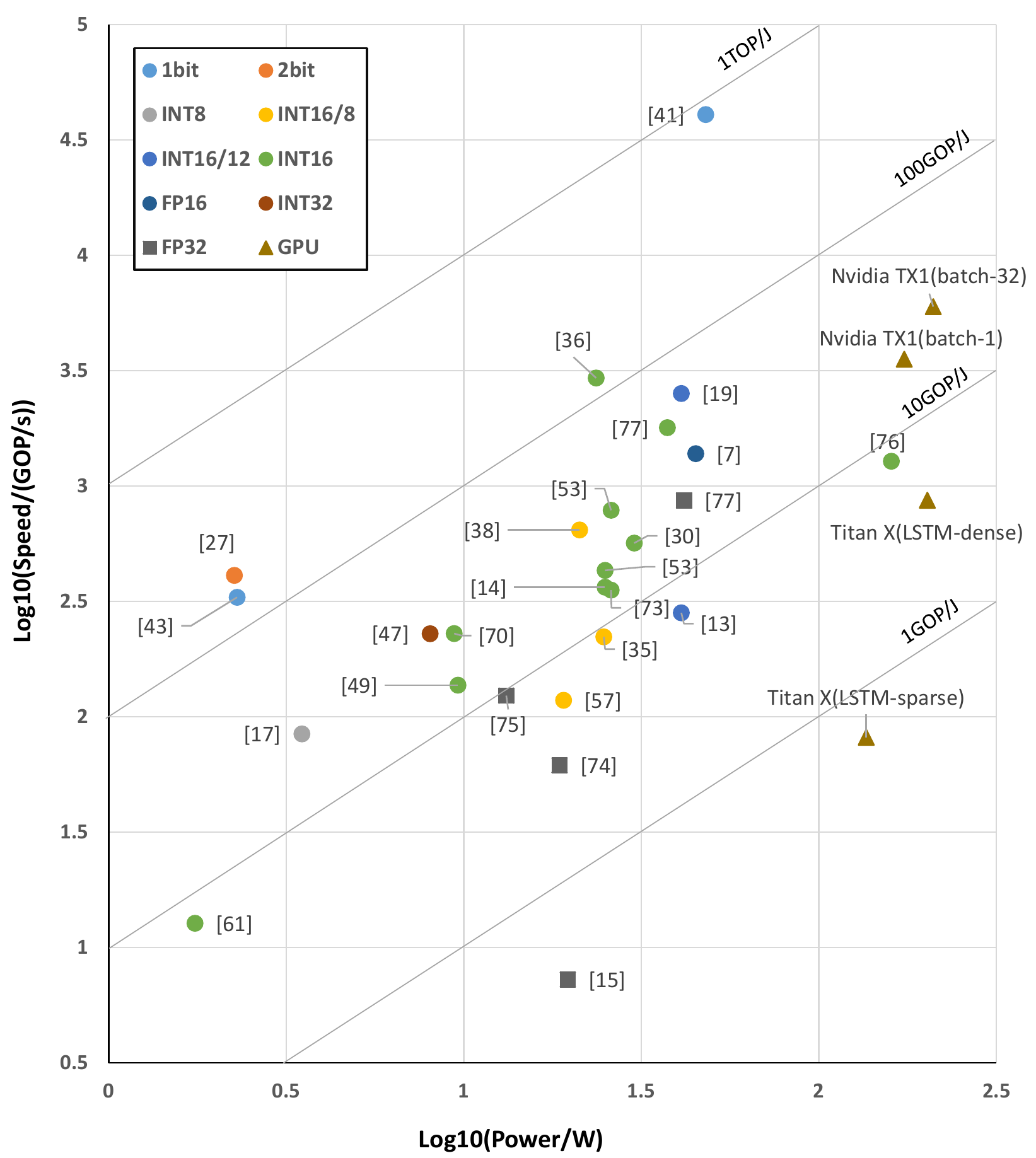}
    \caption{A comparison between different designs on a logarithm coordinate of power and performance. }
    \label{fig:efficiency}
\end{figure}

\begin{figure}[ht]
    \centering
    \includegraphics[width=1.0\columnwidth]{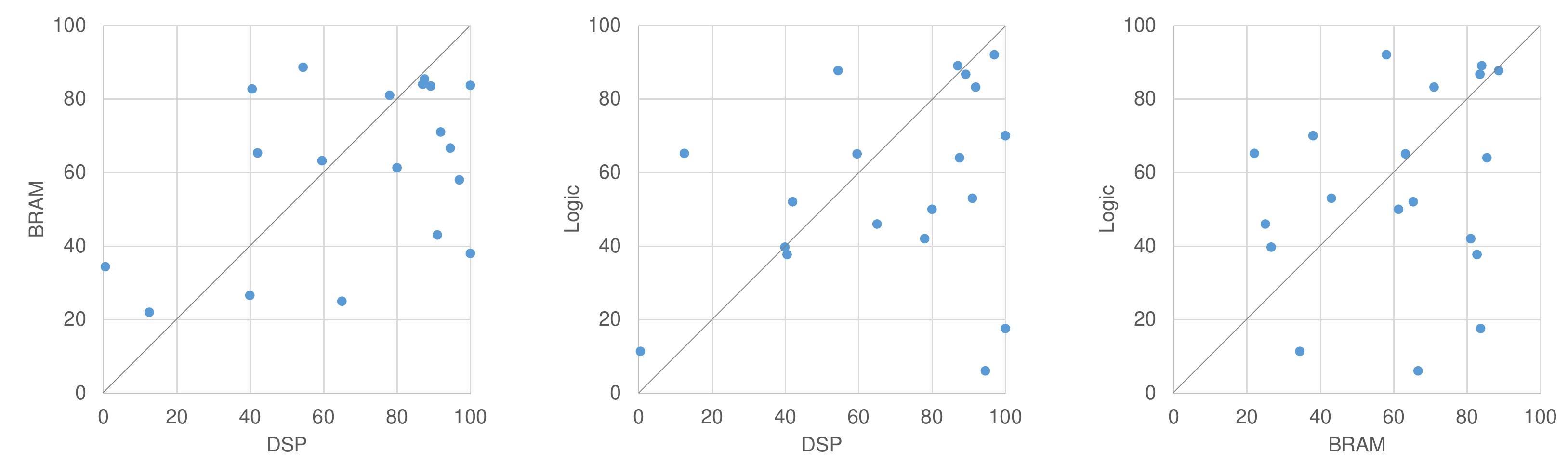}
    \caption{Resource utilization ratio of different accelerator designs.}
    \label{fig:resource}
\end{figure}

Each of the designs in Table~\ref{tab:hardware_list} drawn as a point in Figure~\ref{fig:efficiency}, using $log_{10}(power)$ as $x$ coordinate and $log_{10}(speed)$ as $y$-axis. Therefore, $y-x=log_{10}(energy\_efficiency)$. Besides the FPGA-based designs, we also plot the GPU experimental results used in \cite{guo2017angel, han2017ese} as standards to measure the FPGA designs' performance.

\subsubsection*{\textbf{Bit-width Reduction}} Among all the designs, 1-2 bit based designs~\cite{jiao2017accelerating, moss2017high, nakahara2017fully} show outstanding speed and energy efficiency. This shows that extremely low bit-width is a promising solution for high-performance design. As introduced in section~\ref{sec:software:quant}, linear quantized 1-2 bit network models suffer from great accuracy loss. Further developing related accelerator will be of little use. More efforts should be put on the models. Even trading speed with accuracy can be acceptable considering the current hardware performance.

Besides the 1/2bit designs, the rest of the designs adopts 32-bit floating-point data or linear quantization with 8 or more bits. According to the results in section~\ref{sec:software:quant}, within 1\% accuracy loss can be achieved. So we think the comparison between these designs is fair in accuracy. INT16/8 and INT16 are commonly adopted. But the difference between these designs is not obvious. This is because the underutilization of DSPs discussed in section~\ref{sec:hardware:cu:lbu}. The advantage of INT16 over FP32 is obvious except for \cite{zhang2017improving}, where the hard-core floating-point DSPs are utilized. To a certain extent, this shows the importance of fully utilizing the DSPs on-chip.

\subsubsection*{\textbf{Fast Convolution Algorithm}} Among all the 16-bit designs, \cite{lu2017evaluating} achieves the best energy efficiency and the highest speed with the help of the $6\times 6$ Winograd fast convolution, which is $1.7\times$ faster and $2.6\times$ energy efficient than the 16-bit design in \cite{zhang2017improving}. The design in \cite{zhang2017frequency} achieves $2\times$ speedup and $3\times$ energy efficiency compared with \cite{zhang2015optimizing} where both designs use 32-bit floating-point data and FPGA with 28nm technology node. \rev{Compare with the theoretical $4\times$ performance gain introduced in section~\ref{sec:hardware:cu:fcu}, there is still $1.3-1.5\times$ gap. Not all the layers can use the most optimized fast convolution method because of kernel size limitation.}

\subsubsection*{\textbf{System Level Optimization}} The overall system optimization is not well addressed in most of the work. As this is also related to the HDL design quality, we can roughly evaluate the effect. Here we compare three designs\cite{zhang2016caffeine, liu2016automatic, li2016high} on the same XC7VX690T platform and try to evaluate the effect. All the three designs implement 16-bit fixed-point data format except that ~\cite{liu2016automatic} uses 8-bit for weights. No fast convolution or sparsity is utilized in any of the work. Even though, \cite{li2016high} achieves $2.5\times$ the energy efficiency of \cite{liu2016automatic}. It shows that a system level optimization has a strong effect even comparable to the use of fast convolution algorithm. 

We also investigate the resource utilization of the designs in Table~\ref{tab:hardware_list}. Three kinds of resources (DSP, BRAM, and logic) are considered. We plot the designs in Figure~\ref{fig:resource} using two of the utilization ratio as x and y coordinate. We draw the diagonal line of each figure to show the designs' preference on hardware resource. The BRAM-DSP figure shows an obvious preference on DSP over BRAM. A similar preference appears on DSP over logic. This indicates that current FPGA designs are more likely computation bounded. FPGA manufacturers targeting neural network applications can adjust the resource allocation accordingly. Compared with that, the preference on logic and BRAM seems to be random. A possible explanation is that some of the designers use both logic and DSPs to implement high parallelism, while some prefers to use only DSPs to achieve high working frequency. 

\subsubsection*{\textbf{Comparision with GPU}} In general, FPGA-based designs have achieved comparable energy efficiency to GPU with 10-100GOP/J. But the speed of GPUs still surpasses FPGAs. Scaling up the FPGA-based design is still a problem. Zhang et al.~\cite{zhang2016energy} propose the FPGA-cluster-based solution using 16-bit fixed-point computation. But the energy efficiency is worse than the other 16-bit fixed-point designs. 

Here we estimate the achievable speed of an ideal design. We use the 16-bit fixed-point design in~\cite{lu2017evaluating} as a baseline, which is the best 16-bit fixed-point design with both the highest speed and energy efficiency. 8-bit linear quantization can be adopted according to the analysis in section~\ref{sec:software:quant}, which achieves another $2\times$ speedup and better energy efficiency by utilizing 1 DSP as 2 multipliers. The double frequency optimization further improves the system speed by $2\times$. Consider a sparse model which is similar to the one in~\cite{han2017ese} with 10\% non-zero values. We can estimate a similar $6\times$ improvement as~\cite{han2017ese}. In general about $24\times$ speedup and $12\times$ better energy efficiency can be achieved, which means 72TOP/s speed with about 50W. This shows that it is possible to achieve over $10\times$ higher energy efficiency on FPGA over 32-bit floating-point process on GPU.

The left problem is: does all the techniques: double MAC, sparsification, quantization, fast convolution, and the double frequency design work well together? Pruning a single element in a 2D convolution kernel is of no use for fast convolution because the 2D kernel is always processed as a whole. Directly pruning 2D kernels as a whole may help. But the reported accuracy of this method is lower~\cite{Mao2017Exploring} than a fine-grained pruning. The irregular data access pattern for processing sparse network and the increase in parallelism also brings challenges to the design of memory system and scheduling strategy.

\rev{
\section{Technique Discussion}

To give a better overview of all the techniques introduced in section \ref{sec:software} and \ref{sec:hardware}, we give a brief summary in this section to see how these techniques contribute to FPGA-based NN accelerator designs. Each technique is judged from two aspects: how it affects hardware design and to which level it relates to NN models. Figure~\ref{fig:summary} shows the summary. 

\begin{figure}[ht]
    \centering
    \includegraphics[width=0.8\columnwidth]{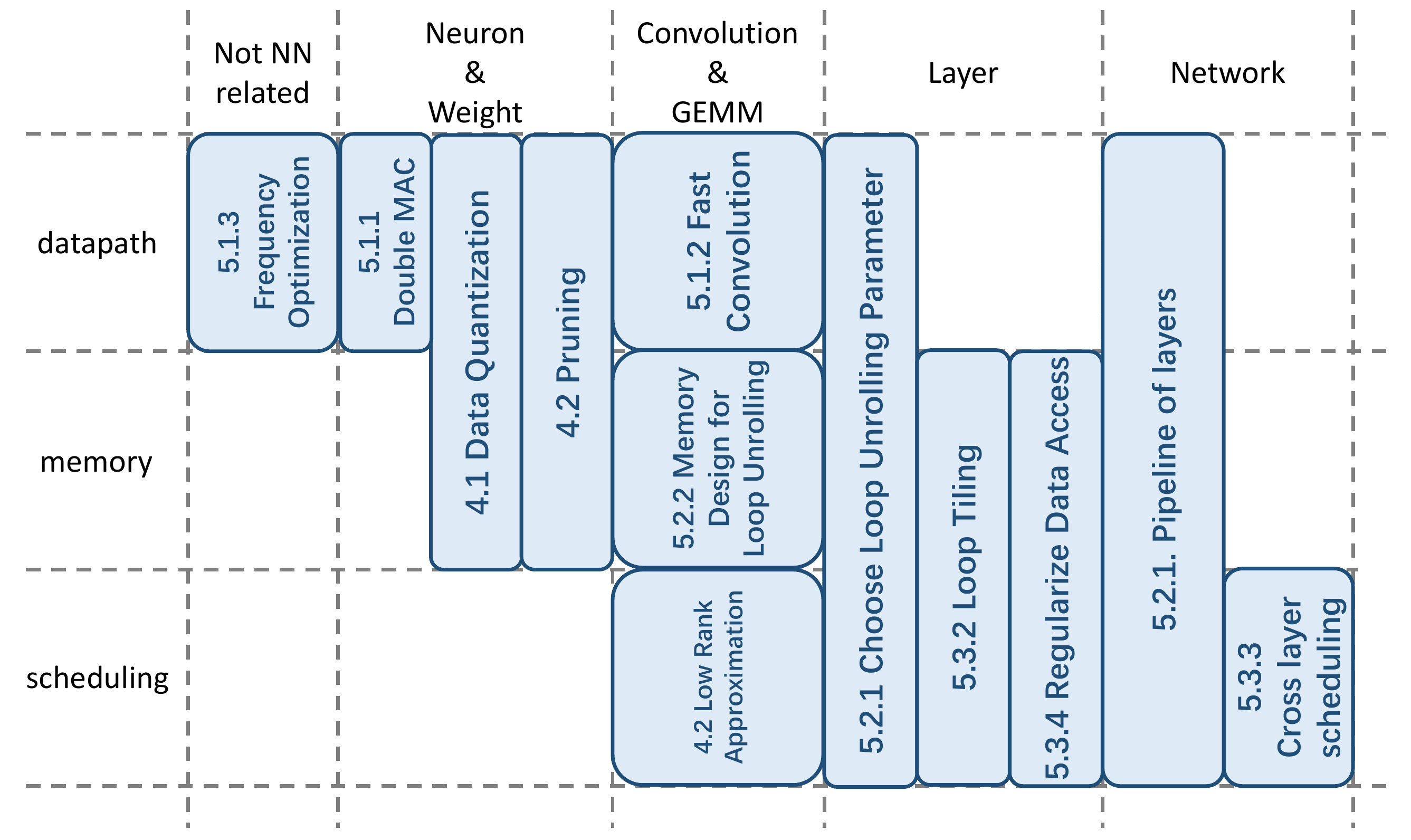}
    \caption{A brief summary of both the software and hardware techniques in section \ref{sec:software} and \ref{sec:hardware}.}
    \label{fig:summary}
\end{figure}

A hardware design basically consists of three parts: datapath, memory, and scheduling. For the design target of high speed, datapath decides the $OPC_{peak}$ while the memory system and scheduling strategy decides $\eta$. For the design target of energy efficiency, datapath decides $E_{op}$ while the memory system decides $N_{SRAM_acc}$ and $N_{DRAM_acc}$. We can see that existing researches are approaching the design target from every aspect by utilizing the neural network model features from single neuron level to the whole network level.

What is the future of FPGA-based neural network inference accelerator? Currently, much of the techniques lie in the neuron level and the convolution level. There are two reasons for this. The first reason is that few feature can be utilized in layer level and network level. Most of the existing NN models introduce a simple structure with cascaded layers~\cite{krizhevsky2012imagenet,simonyan2014very} or simply adding a by-path~\cite{he2016deep}. New features like depth-wise convolution~\cite{Howard2017MobileNets} and the complex branch in SSD~\cite{liu2016ssd} may brings more design opportunities. But few work focuses on these models. The second reason is that the scale of an FPGA chip is limited. An FPGA chip usually consists of hundreds to thousands of DSPs. This number is still too small compared with a single neural network layer with more than 100M operations. 

So further opportunities may come from two aspects. The first is the evolution of network structure. The second is the scaling up of FPGA-based system, with larger chips or multiple chips. Existing designs using small models with binary weights are making their FPGAs relatively larger. These designs already introduce some subversive ideas like mapping the whole networks spatially onto hardware~\cite{yang2018fully}. Besides the opportunities, designers are also faced with the scaling up challenges, from the limitation of loop unrolling, bandwidth, etc. 
}

\section{Design Automation and Flexibility}\label{sec:flexibility}

Mapping a certain CNN model onto an FPGA accelerator still requires much heavier work than developing with existing deep learning frameworks. In some application scenarios, various NN models are to be supported with the FPGA accelerator. Thus the design automation of CNN accelerators is also important. Various researches have been focusing on CNN accelerator design toolflows. Venieris, et al.~\cite{venieris2018toolflows} give a detailed discussion on different toolflows in supported models, interface, hardware architecture, design space exploration and arithmetic precision. In this chapter, as we have been focusing on detailed techniques used in model optimization and hardware design, we only classify the toolflows into two categories: hardware design automation and software design automation. Hardware design automation generates different hardware designs according to different NN models. Software design automation keeps the same accelerator and generates different inputs to the accelerator. The discussion in this section can serve as a supplementary to ~\cite{venieris2018toolflows}. 

\subsection{Hardware Design Automation}
Hardware design automation is widely adopted in FPGA-based accelerators because of the reconfigurability of FPGAs~\cite{venieris2017fpgaconvnet, morcel2017minimalist, ma2017automatic, venieris2017latency, dicecco2016caffeinated, wang2016deepburning, sharma2016high}. These proposed techniques focus on automatically generate the HDL design based on the network parameter. Difference between these methods is the selection of an intermediate level description of the network to cover the gap between high-level network description and low-level hardware design.

A straightforward way is no intermediate description. The design flow in \cite{ma2017automatic} searches the optimized parameter for a handcrafted Verilog template with the input network description and platform constraint. This method is similar to the optimization methods mentioned in section~\ref{sec:hardware}. DiCecco et al.~\cite{dicecco2016caffeinated} use a similar idea based on OpenCL model. This enables that the development tool be integrated with Caffe and one network can be executed on different platforms. 

Venireis, et al.~\cite{venieris2017latency} describes the network model as a DFG in their design tool. Then the network computaion process is translated to hardware design with DFG mapping method.

DnnWeaver~\cite{sharma2016high} use a virtual instruction set to describe a network. The network model is first translated into an instruction sequence. Then the sequence is mapped as hardware FSM states but not executed like traditional CPU instructions. 

Hardware design automation directly modifies the hardware design to support different networks. This means the hardware can always achieve the best performance on the target platform. This is suitable for FPGA because of its reconfigurability. It works in situations where network switching is not frequent and the reconfiguration overhead does not care. For example, for a large-scale cloud service, the change in network models can be covered by switching between different FPGA chips. So the FPGAs do not need to be reconfigured frequently.

\subsection{Software Design Automation}

Software design automation tries to run different networks on the same hardware accelerator by simply changing the input, in most cases, an instruction sequence. The difference between these work is the granularity of instruction. At a lower level, Guo, et al.~\cite{guo2017angel} propose the instruction set with only three kinds of instructions: LOAD, CALC, and SAVE. The granularity of the LOAD and SAVE instructions are the same as the data tiling size. Each CONV executes a set of 2-D convolutions given the feature map size encoded in the instruction. The channel number is fixed as the hardware unrolling parameter. At this level, the software compiler can carry out static scheduling and dynamic data reuse strategy accordingly for each layer. DNNDK~\cite{dnndk} uses similar ideas but with more functions in the instructions to support various networks.

Zhang et al.~\cite{zhang2016caffeine} use a layer level instruction set. The control of a CNN layer is designed as a configurable hardware FSM. Compared with \cite{guo2017angel}, this reduces the memory access for instruction while increasing the hardware cost on the configurable FSM.

TVM~\cite{chen2018tvm} implements a uniform mapping optimization framework for different kinds of platforms including CPU, GPU, FPGA, and ASIC. The framework allows developers to define customized parallel primitive to support customized hardware, including FPGA accelerators. This means the scheduling granularity is more flexible.

Instruction based methods do not modify hardware and thus enables that the accelerator can switch between networks at run-time. An example of the application scenario is the real-time video processing system on a mobile platform. The process of a single frame can involve different networks if the task is complex enough. Reconfigure the hardware causes unacceptable overhead while instruction based methods can solve the problem if all the instructions of all the networks are prepared in memory.

\subsection{Mixed Method}
Wang, et al.~\cite{wang2016deepburning} propose a design automation framework mixing the above two by both optimizing hardware design and compile software instructions. The hardware is first assembled with pre-defined HDL templates using the optimized hardware parameter. The data control flow of the computation process is controlled by software binaries, which is compiled according to the network description. It is possible that the hardware can be used for a new network by simply changing the software binaries.

\section{Conclusion}\label{sec:conclusion}

In this paper, we review state-of-the-art neural network accelerator designs and summarize the techniques used. According to the evaluation result in section~\ref{sec:evaluation}, with software hardware co-design, FPGA can achieve more than $10\times$ better speed and energy efficiency than state-of-the-art GPU. This shows that FPGA is a promising candidate for neural network acceleration. We also review the methods used for accelerator design automation, which shows that current development flow can achieve both high performance and run-time network switch.

But there is still a gap between current designs and the estimation. On the one hand, quantization with extremely narrow bit-width is limited by the model accuracy, which needs further algorithm research. On the other hand, combining all the techniques needs more research in both software and hardware to make them work well together. Commercial tools including DNNDK~\cite{dnndk} is taking a first step but still has a lone way to go. Scaling up the design is also a problem. Future work should focus on solving these challenges. 

\section*{Acknowledgement}
This work was supported by National Key R\&D Program of China (2018YFB0105005, 2017YFA0207600), National Natural Science Foundation of China (No. 61622403, 61621091), DeePhi Technology and Xilinx.

\bibliographystyle{ACM-Reference-Format}
\bibliography{ref}


\begin{thebibliography}{84}


\ifx \showCODEN    \undefined \def \showCODEN     #1{\unskip}     \fi
\ifx \showDOI      \undefined \def \showDOI       #1{#1}\fi
\ifx \showISBNx    \undefined \def \showISBNx     #1{\unskip}     \fi
\ifx \showISBNxiii \undefined \def \showISBNxiii  #1{\unskip}     \fi
\ifx \showISSN     \undefined \def \showISSN      #1{\unskip}     \fi
\ifx \showLCCN     \undefined \def \showLCCN      #1{\unskip}     \fi
\ifx \shownote     \undefined \def \shownote      #1{#1}          \fi
\ifx \showarticletitle \undefined \def \showarticletitle #1{#1}   \fi
\ifx \showURL      \undefined \def \showURL       {\relax}        \fi
\providecommand\bibfield[2]{#2}
\providecommand\bibinfo[2]{#2}
\providecommand\natexlab[1]{#1}
\providecommand\showeprint[2][]{arXiv:#2}

\bibitem[\protect\citeauthoryear{??}{cha}{[n. d.]}]%
        {chai_dnn}
 \bibinfo{year}{[n. d.]}\natexlab{}.
\newblock \bibinfo{howpublished}{\url{https://github.com/Xilinx/chaidnn}}.
  (\bibinfo{year}{[n. d.]}).
\newblock
\newblock
\shownote{Accessed August 23, 2018.}


\bibitem[\protect\citeauthoryear{??}{xfd}{[n. d.]}]%
        {xfdnn}
 \bibinfo{year}{[n. d.]}\natexlab{}.
\newblock
  \bibinfo{howpublished}{\url{https://www.xilinx.com/support/documentation/white_papers/wp504-accel-dnns.pdf}}.
    (\bibinfo{year}{[n. d.]}).
\newblock
\newblock
\shownote{Accessed December 3, 2018.}


\bibitem[\protect\citeauthoryear{??}{dnn}{[n. d.]}]%
        {dnndk}
 \bibinfo{year}{[n. d.]}\natexlab{}.
\newblock \bibinfo{howpublished}{\url{http://www.deephi.com/technology/dnndk}}.
    (\bibinfo{year}{[n. d.]}).
\newblock
\newblock
\shownote{Accessed December 3, 2018.}


\bibitem[\protect\citeauthoryear{Abadi, Agarwal, Barham, Brevdo, Chen, Citro,
  Corrado, Davis, Dean, Devin, et~al\mbox{.}}{Abadi et~al\mbox{.}}{2016}]%
        {abadi2016tensorflow}
\bibfield{author}{\bibinfo{person}{Mart{\'\i}n Abadi}, \bibinfo{person}{Ashish
  Agarwal}, \bibinfo{person}{Paul Barham}, \bibinfo{person}{Eugene Brevdo},
  \bibinfo{person}{Zhifeng Chen}, \bibinfo{person}{Craig Citro},
  \bibinfo{person}{Greg~S Corrado}, \bibinfo{person}{Andy Davis},
  \bibinfo{person}{Jeffrey Dean}, \bibinfo{person}{Matthieu Devin},
  {et~al\mbox{.}}} \bibinfo{year}{2016}\natexlab{}.
\newblock \showarticletitle{Tensorflow: Large-scale machine learning on
  heterogeneous distributed systems}.
\newblock \bibinfo{journal}{{\em arXiv preprint arXiv:1603.04467\/}}
  (\bibinfo{year}{2016}).
\newblock


\bibitem[\protect\citeauthoryear{Alwani, Chen, Ferdman, and Milder}{Alwani
  et~al\mbox{.}}{2016}]%
        {alwani2016fused}
\bibfield{author}{\bibinfo{person}{Manoj Alwani}, \bibinfo{person}{Han Chen},
  \bibinfo{person}{Michael Ferdman}, {and} \bibinfo{person}{Peter Milder}.}
  \bibinfo{year}{2016}\natexlab{}.
\newblock \showarticletitle{Fused-layer CNN accelerators}. In
  \bibinfo{booktitle}{{\em Microarchitecture (MICRO), 2016 49th Annual IEEE/ACM
  International Symposium on}}. IEEE, \bibinfo{pages}{1--12}.
\newblock


\bibitem[\protect\citeauthoryear{Amodei, Ananthanarayanan, Anubhai, Bai,
  Battenberg, Case, Casper, Catanzaro, Cheng, Chen, et~al\mbox{.}}{Amodei
  et~al\mbox{.}}{2016}]%
        {amodei2016deep}
\bibfield{author}{\bibinfo{person}{Dario Amodei}, \bibinfo{person}{Sundaram
  Ananthanarayanan}, \bibinfo{person}{Rishita Anubhai},
  \bibinfo{person}{Jingliang Bai}, \bibinfo{person}{Eric Battenberg},
  \bibinfo{person}{Carl Case}, \bibinfo{person}{Jared Casper},
  \bibinfo{person}{Bryan Catanzaro}, \bibinfo{person}{Qiang Cheng},
  \bibinfo{person}{Guoliang Chen}, {et~al\mbox{.}}}
  \bibinfo{year}{2016}\natexlab{}.
\newblock \showarticletitle{Deep speech 2: End-to-end speech recognition in
  english and mandarin}. In \bibinfo{booktitle}{{\em International Conference
  on Machine Learning}}. \bibinfo{pages}{173--182}.
\newblock


\bibitem[\protect\citeauthoryear{Aydonat, O'Connell, Capalija, Ling, and
  Chiu}{Aydonat et~al\mbox{.}}{2017}]%
        {aydonat2017opencl}
\bibfield{author}{\bibinfo{person}{Utku Aydonat}, \bibinfo{person}{Shane
  O'Connell}, \bibinfo{person}{Davor Capalija}, \bibinfo{person}{Andrew~C
  Ling}, {and} \bibinfo{person}{Gordon~R Chiu}.}
  \bibinfo{year}{2017}\natexlab{}.
\newblock \showarticletitle{An OpenCL (TM) Deep Learning Accelerator on Arria
  10}.
\newblock \bibinfo{journal}{{\em arXiv preprint arXiv:1701.03534\/}}
  (\bibinfo{year}{2017}).
\newblock


\bibitem[\protect\citeauthoryear{Chen, Moreau, Jiang, Zheng, Yan, Shen, Cowan,
  Wang, Hu, Ceze, et~al\mbox{.}}{Chen et~al\mbox{.}}{2018}]%
        {chen2018tvm}
\bibfield{author}{\bibinfo{person}{Tianqi Chen}, \bibinfo{person}{Thierry
  Moreau}, \bibinfo{person}{Ziheng Jiang}, \bibinfo{person}{Lianmin Zheng},
  \bibinfo{person}{Eddie Yan}, \bibinfo{person}{Haichen Shen},
  \bibinfo{person}{Meghan Cowan}, \bibinfo{person}{Leyuan Wang},
  \bibinfo{person}{Yuwei Hu}, \bibinfo{person}{Luis Ceze}, {et~al\mbox{.}}}
  \bibinfo{year}{2018}\natexlab{}.
\newblock \showarticletitle{$\{$TVM$\}$: An Automated End-to-End Optimizing
  Compiler for Deep Learning}. In \bibinfo{booktitle}{{\em 13th $\{$USENIX$\}$
  Symposium on Operating Systems Design and Implementation ($\{$OSDI$\}$ 18)}}.
  \bibinfo{pages}{578--594}.
\newblock


\bibitem[\protect\citeauthoryear{Chen, Wilson, Tyree, Weinberger, and
  Chen}{Chen et~al\mbox{.}}{2015}]%
        {chen2015compressing}
\bibfield{author}{\bibinfo{person}{Wenlin Chen}, \bibinfo{person}{James
  Wilson}, \bibinfo{person}{Stephen Tyree}, \bibinfo{person}{Kilian
  Weinberger}, {and} \bibinfo{person}{Yixin Chen}.}
  \bibinfo{year}{2015}\natexlab{}.
\newblock \showarticletitle{Compressing neural networks with the hashing
  trick}. In \bibinfo{booktitle}{{\em International Conference on Machine
  Learning}}. \bibinfo{pages}{2285--2294}.
\newblock


\bibitem[\protect\citeauthoryear{DiCecco, Lacey, Vasiljevic, Chow, Taylor, and
  Areibi}{DiCecco et~al\mbox{.}}{2016}]%
        {dicecco2016caffeinated}
\bibfield{author}{\bibinfo{person}{Roberto DiCecco}, \bibinfo{person}{Griffin
  Lacey}, \bibinfo{person}{Jasmina Vasiljevic}, \bibinfo{person}{Paul Chow},
  \bibinfo{person}{Graham Taylor}, {and} \bibinfo{person}{Shawki Areibi}.}
  \bibinfo{year}{2016}\natexlab{}.
\newblock \showarticletitle{Caffeinated FPGAs: FPGA Framework For Convolutional
  Neural Networks}. In \bibinfo{booktitle}{{\em Field-Programmable Technology
  (FPT), 2016 International Conference on}}. IEEE, \bibinfo{pages}{265--268}.
\newblock


\bibitem[\protect\citeauthoryear{Ding, Liao, Wang, Li, Liu, Zhuo, Wang, Qian,
  Bai, Yuan, et~al\mbox{.}}{Ding et~al\mbox{.}}{2017}]%
        {ding2017c}
\bibfield{author}{\bibinfo{person}{Caiwen Ding}, \bibinfo{person}{Siyu Liao},
  \bibinfo{person}{Yanzhi Wang}, \bibinfo{person}{Zhe Li},
  \bibinfo{person}{Ning Liu}, \bibinfo{person}{Youwei Zhuo},
  \bibinfo{person}{Chao Wang}, \bibinfo{person}{Xuehai Qian},
  \bibinfo{person}{Yu Bai}, \bibinfo{person}{Geng Yuan}, {et~al\mbox{.}}}
  \bibinfo{year}{2017}\natexlab{}.
\newblock \showarticletitle{CirCNN: accelerating and compressing deep neural
  networks using block-circulant weight matrices}. In \bibinfo{booktitle}{{\em
  Proceedings of the 50th Annual IEEE/ACM International Symposium on
  Microarchitecture}}. ACM, \bibinfo{pages}{395--408}.
\newblock


\bibitem[\protect\citeauthoryear{Ghasemzadeh, Samragh, and
  Koushanfar}{Ghasemzadeh et~al\mbox{.}}{[n. d.]}]%
        {ghasemzadehrebnet}
\bibfield{author}{\bibinfo{person}{Mohammad Ghasemzadeh},
  \bibinfo{person}{Mohammad Samragh}, {and} \bibinfo{person}{Farinaz
  Koushanfar}.} \bibinfo{year}{[n. d.]}\natexlab{}.
\newblock \showarticletitle{ReBNet: Residual Binarized Neural Network}.
\newblock  (\bibinfo{year}{[n. d.]}).
\newblock


\bibitem[\protect\citeauthoryear{Girshick, Donahue, Darrell, and
  Malik}{Girshick et~al\mbox{.}}{2014}]%
        {girshick2014rich}
\bibfield{author}{\bibinfo{person}{Ross Girshick}, \bibinfo{person}{Jeff
  Donahue}, \bibinfo{person}{Trevor Darrell}, {and} \bibinfo{person}{Jitendra
  Malik}.} \bibinfo{year}{2014}\natexlab{}.
\newblock \showarticletitle{Rich feature hierarchies for accurate object
  detection and semantic segmentation}. In \bibinfo{booktitle}{{\em Proceedings
  of the IEEE conference on computer vision and pattern recognition}}.
  \bibinfo{pages}{580--587}.
\newblock


\bibitem[\protect\citeauthoryear{Guan, Liang, Xu, Wang, Shi, Chen, Sun, Zhang,
  and Cong}{Guan et~al\mbox{.}}{2017a}]%
        {guan2017fp}
\bibfield{author}{\bibinfo{person}{Yijin Guan}, \bibinfo{person}{Hao Liang},
  \bibinfo{person}{Ningyi Xu}, \bibinfo{person}{Wenqiang Wang},
  \bibinfo{person}{Shaoshuai Shi}, \bibinfo{person}{Xi Chen},
  \bibinfo{person}{Guangyu Sun}, \bibinfo{person}{Wei Zhang}, {and}
  \bibinfo{person}{Jason Cong}.} \bibinfo{year}{2017}\natexlab{a}.
\newblock \showarticletitle{FP-DNN: An Automated Framework for Mapping Deep
  Neural Networks onto FPGAs with RTL-HLS Hybrid Templates}. In
  \bibinfo{booktitle}{{\em Field-Programmable Custom Computing Machines (FCCM),
  2017 IEEE 25th Annual International Symposium on}}. IEEE,
  \bibinfo{pages}{152--159}.
\newblock


\bibitem[\protect\citeauthoryear{Guan, Yuan, Sun, and Cong}{Guan
  et~al\mbox{.}}{2017b}]%
        {guan2017fpga}
\bibfield{author}{\bibinfo{person}{Yijin Guan}, \bibinfo{person}{Zhihang Yuan},
  \bibinfo{person}{Guangyu Sun}, {and} \bibinfo{person}{Jason Cong}.}
  \bibinfo{year}{2017}\natexlab{b}.
\newblock \showarticletitle{FPGA-based accelerator for long short-term memory
  recurrent neural networks}. In \bibinfo{booktitle}{{\em Design Automation
  Conference (ASP-DAC), 2017 22nd Asia and South Pacific}}. IEEE,
  \bibinfo{pages}{629--634}.
\newblock


\bibitem[\protect\citeauthoryear{Guo, Yin, Ouyang, Liu, and Wei}{Guo
  et~al\mbox{.}}{2017b}]%
        {guo2017bit}
\bibfield{author}{\bibinfo{person}{Jianxin Guo}, \bibinfo{person}{Shouyi Yin},
  \bibinfo{person}{Peng Ouyang}, \bibinfo{person}{Leibo Liu}, {and}
  \bibinfo{person}{Shaojun Wei}.} \bibinfo{year}{2017}\natexlab{b}.
\newblock \showarticletitle{Bit-Width Based Resource Partitioning for CNN
  Acceleration on FPGA}. In \bibinfo{booktitle}{{\em Field-Programmable Custom
  Computing Machines (FCCM), 2017 IEEE 25th Annual International Symposium
  on}}. IEEE, \bibinfo{pages}{31--31}.
\newblock


\bibitem[\protect\citeauthoryear{Guo, Sui, Qiu, Yu, Wang, Yao, Han, Wang, and
  Yang}{Guo et~al\mbox{.}}{2017a}]%
        {guo2017angel}
\bibfield{author}{\bibinfo{person}{Kaiyuan Guo}, \bibinfo{person}{Lingzhi Sui},
  \bibinfo{person}{Jiantao Qiu}, \bibinfo{person}{Jincheng Yu},
  \bibinfo{person}{Junbin Wang}, \bibinfo{person}{Song Yao},
  \bibinfo{person}{Song Han}, \bibinfo{person}{Yu Wang}, {and}
  \bibinfo{person}{Huazhong Yang}.} \bibinfo{year}{2017}\natexlab{a}.
\newblock \showarticletitle{Angel-Eye: A Complete Design Flow for Mapping CNN
  onto Embedded FPGA}.
\newblock \bibinfo{journal}{{\em IEEE Transactions on Computer-Aided Design of
  Integrated Circuits and Systems\/}} (\bibinfo{year}{2017}).
\newblock


\bibitem[\protect\citeauthoryear{Gupta}{Gupta}{2016}]%
        {gupta2016accelerating}
\bibfield{author}{\bibinfo{person}{PK Gupta}.} \bibinfo{year}{2016}\natexlab{}.
\newblock \showarticletitle{Accelerating datacenter workloads}. In
  \bibinfo{booktitle}{{\em 26th International Conference on Field Programmable
  Logic and Applications (FPL)}}.
\newblock


\bibitem[\protect\citeauthoryear{Han, Kang, Mao, Hu, Li, Li, Xie, Luo, Yao,
  Wang, et~al\mbox{.}}{Han et~al\mbox{.}}{2017}]%
        {han2017ese}
\bibfield{author}{\bibinfo{person}{Song Han}, \bibinfo{person}{Junlong Kang},
  \bibinfo{person}{Huizi Mao}, \bibinfo{person}{Yiming Hu},
  \bibinfo{person}{Xin Li}, \bibinfo{person}{Yubin Li},
  \bibinfo{person}{Dongliang Xie}, \bibinfo{person}{Hong Luo},
  \bibinfo{person}{Song Yao}, \bibinfo{person}{Yu Wang}, {et~al\mbox{.}}}
  \bibinfo{year}{2017}\natexlab{}.
\newblock \showarticletitle{ESE: Efficient Speech Recognition Engine with
  Sparse LSTM on FPGA.}. In \bibinfo{booktitle}{{\em FPGA}}.
  \bibinfo{pages}{75--84}.
\newblock


\bibitem[\protect\citeauthoryear{Han, Mao, and Dally}{Han
  et~al\mbox{.}}{2015}]%
        {han2015deep}
\bibfield{author}{\bibinfo{person}{Song Han}, \bibinfo{person}{Huizi Mao},
  {and} \bibinfo{person}{William~J Dally}.} \bibinfo{year}{2015}\natexlab{}.
\newblock \showarticletitle{Deep compression: Compressing deep neural networks
  with pruning, trained quantization and huffman coding}.
\newblock \bibinfo{journal}{{\em arXiv preprint arXiv:1510.00149\/}}
  (\bibinfo{year}{2015}).
\newblock


\bibitem[\protect\citeauthoryear{Hannun, Case, Casper, Catanzaro, Diamos,
  Elsen, Prenger, Satheesh, Sengupta, Coates, et~al\mbox{.}}{Hannun
  et~al\mbox{.}}{2014}]%
        {hannun2014deep}
\bibfield{author}{\bibinfo{person}{Awni Hannun}, \bibinfo{person}{Carl Case},
  \bibinfo{person}{Jared Casper}, \bibinfo{person}{Bryan Catanzaro},
  \bibinfo{person}{Greg Diamos}, \bibinfo{person}{Erich Elsen},
  \bibinfo{person}{Ryan Prenger}, \bibinfo{person}{Sanjeev Satheesh},
  \bibinfo{person}{Shubho Sengupta}, \bibinfo{person}{Adam Coates},
  {et~al\mbox{.}}} \bibinfo{year}{2014}\natexlab{}.
\newblock \showarticletitle{Deep speech: Scaling up end-to-end speech
  recognition}.
\newblock \bibinfo{journal}{{\em arXiv preprint arXiv:1412.5567\/}}
  (\bibinfo{year}{2014}).
\newblock


\bibitem[\protect\citeauthoryear{He, Zhang, Ren, and Sun}{He
  et~al\mbox{.}}{2016}]%
        {he2016deep}
\bibfield{author}{\bibinfo{person}{Kaiming He}, \bibinfo{person}{Xiangyu
  Zhang}, \bibinfo{person}{Shaoqing Ren}, {and} \bibinfo{person}{Jian Sun}.}
  \bibinfo{year}{2016}\natexlab{}.
\newblock \showarticletitle{Deep residual learning for image recognition}. In
  \bibinfo{booktitle}{{\em Proceedings of the IEEE conference on computer
  vision and pattern recognition}}. \bibinfo{pages}{770--778}.
\newblock


\bibitem[\protect\citeauthoryear{Horowitz}{Horowitz}{[n. d.]}]%
        {vlsi_energy}
\bibfield{author}{\bibinfo{person}{M. Horowitz}.} \bibinfo{year}{[n.
  d.]}\natexlab{}.
\newblock \bibinfo{title}{Energy table for 45nm process, Stanford VLSI
  wiki.[Online].}
\newblock
  \bibinfo{howpublished}{\url{https://sites.google.com/site/seecproject}}.
  (\bibinfo{year}{[n. d.]}).
\newblock


\bibitem[\protect\citeauthoryear{Howard, Zhu, Chen, Kalenichenko, Wang, Weyand,
  Andreetto, and Adam}{Howard et~al\mbox{.}}{2017}]%
        {Howard2017MobileNets}
\bibfield{author}{\bibinfo{person}{Andrew~G Howard}, \bibinfo{person}{Menglong
  Zhu}, \bibinfo{person}{Bo Chen}, \bibinfo{person}{Dmitry Kalenichenko},
  \bibinfo{person}{Weijun Wang}, \bibinfo{person}{Tobias Weyand},
  \bibinfo{person}{Marco Andreetto}, {and} \bibinfo{person}{Hartwig Adam}.}
  \bibinfo{year}{2017}\natexlab{}.
\newblock \showarticletitle{MobileNets: Efficient Convolutional Neural Networks
  for Mobile Vision Applications}.
\newblock  (\bibinfo{year}{2017}).
\newblock


\bibitem[\protect\citeauthoryear{Iandola, Han, Moskewicz, Ashraf, Dally, and
  Keutzer}{Iandola et~al\mbox{.}}{2016}]%
        {iandola2016squeezenet}
\bibfield{author}{\bibinfo{person}{Forrest~N Iandola}, \bibinfo{person}{Song
  Han}, \bibinfo{person}{Matthew~W Moskewicz}, \bibinfo{person}{Khalid Ashraf},
  \bibinfo{person}{William~J Dally}, {and} \bibinfo{person}{Kurt Keutzer}.}
  \bibinfo{year}{2016}\natexlab{}.
\newblock \showarticletitle{SqueezeNet: AlexNet-level accuracy with 50x fewer
  parameters and< 0.5 MB model size}.
\newblock \bibinfo{journal}{{\em arXiv preprint arXiv:1602.07360\/}}
  (\bibinfo{year}{2016}).
\newblock


\bibitem[\protect\citeauthoryear{Jia, Shelhamer, Donahue, Karayev, Long,
  Girshick, Guadarrama, and Darrell}{Jia et~al\mbox{.}}{2014}]%
        {jia2014caffe}
\bibfield{author}{\bibinfo{person}{Yangqing Jia}, \bibinfo{person}{Evan
  Shelhamer}, \bibinfo{person}{Jeff Donahue}, \bibinfo{person}{Sergey Karayev},
  \bibinfo{person}{Jonathan Long}, \bibinfo{person}{Ross Girshick},
  \bibinfo{person}{Sergio Guadarrama}, {and} \bibinfo{person}{Trevor Darrell}.}
  \bibinfo{year}{2014}\natexlab{}.
\newblock \showarticletitle{Caffe: Convolutional Architecture for Fast Feature
  Embedding}.
\newblock \bibinfo{journal}{{\em arXiv preprint arXiv:1408.5093\/}}
  (\bibinfo{year}{2014}).
\newblock


\bibitem[\protect\citeauthoryear{Jiao, Luo, Cao, Zhou, and Wang}{Jiao
  et~al\mbox{.}}{2017}]%
        {jiao2017accelerating}
\bibfield{author}{\bibinfo{person}{Li Jiao}, \bibinfo{person}{Cheng Luo},
  \bibinfo{person}{Wei Cao}, \bibinfo{person}{Xuegong Zhou}, {and}
  \bibinfo{person}{Lingli Wang}.} \bibinfo{year}{2017}\natexlab{}.
\newblock \showarticletitle{Accelerating low bit-width convolutional neural
  networks with embedded FPGA}. In \bibinfo{booktitle}{{\em Field Programmable
  Logic and Applications (FPL), 2017 27th International Conference on}}. IEEE,
  \bibinfo{pages}{1--4}.
\newblock


\bibitem[\protect\citeauthoryear{Krizhevsky, Sutskever, and Hinton}{Krizhevsky
  et~al\mbox{.}}{2012}]%
        {krizhevsky2012imagenet}
\bibfield{author}{\bibinfo{person}{Alex Krizhevsky}, \bibinfo{person}{Ilya
  Sutskever}, {and} \bibinfo{person}{Geoffrey~E Hinton}.}
  \bibinfo{year}{2012}\natexlab{}.
\newblock \showarticletitle{Imagenet classification with deep convolutional
  neural networks}. In \bibinfo{booktitle}{{\em Advances in neural information
  processing systems}}. \bibinfo{pages}{1097--1105}.
\newblock


\bibitem[\protect\citeauthoryear{Li, Zhang, and Liu}{Li et~al\mbox{.}}{2016b}]%
        {li2016ternary}
\bibfield{author}{\bibinfo{person}{Fengfu Li}, \bibinfo{person}{Bo Zhang},
  {and} \bibinfo{person}{Bin Liu}.} \bibinfo{year}{2016}\natexlab{b}.
\newblock \showarticletitle{Ternary weight networks}.
\newblock \bibinfo{journal}{{\em arXiv preprint arXiv:1605.04711\/}}
  (\bibinfo{year}{2016}).
\newblock


\bibitem[\protect\citeauthoryear{Li, Fan, Jiao, Cao, Zhou, and Wang}{Li
  et~al\mbox{.}}{2016a}]%
        {li2016high}
\bibfield{author}{\bibinfo{person}{Huimin Li}, \bibinfo{person}{Xitian Fan},
  \bibinfo{person}{Li Jiao}, \bibinfo{person}{Wei Cao},
  \bibinfo{person}{Xuegong Zhou}, {and} \bibinfo{person}{Lingli Wang}.}
  \bibinfo{year}{2016}\natexlab{a}.
\newblock \showarticletitle{A high performance FPGA-based accelerator for
  large-scale convolutional neural networks}. In \bibinfo{booktitle}{{\em Field
  Programmable Logic and Applications (FPL), 2016 26th International Conference
  on}}. IEEE, \bibinfo{pages}{1--9}.
\newblock


\bibitem[\protect\citeauthoryear{Li, Liu, Xu, Yu, and Ren}{Li
  et~al\mbox{.}}{2017}]%
        {li20177}
\bibfield{author}{\bibinfo{person}{Yixing Li}, \bibinfo{person}{Zichuan Liu},
  \bibinfo{person}{Kai Xu}, \bibinfo{person}{Hao Yu}, {and}
  \bibinfo{person}{Fengbo Ren}.} \bibinfo{year}{2017}\natexlab{}.
\newblock \showarticletitle{A 7.663-TOPS 8.2-W Energy-efficient FPGA
  Accelerator for Binary Convolutional Neural Networks}.
\newblock \bibinfo{journal}{{\em arXiv preprint arXiv:1702.06392\/}}
  (\bibinfo{year}{2017}).
\newblock


\bibitem[\protect\citeauthoryear{Lin, Yin, Tu, Liu, Li, and Wei}{Lin
  et~al\mbox{.}}{2018}]%
        {lin2018lcp}
\bibfield{author}{\bibinfo{person}{Xinhan Lin}, \bibinfo{person}{Shouyi Yin},
  \bibinfo{person}{Fengbin Tu}, \bibinfo{person}{Leibo Liu},
  \bibinfo{person}{Xiangyu Li}, {and} \bibinfo{person}{Shaojun Wei}.}
  \bibinfo{year}{2018}\natexlab{}.
\newblock \showarticletitle{LCP: a layer clusters paralleling mapping method
  for accelerating inception and residual networks on FPGA}. In
  \bibinfo{booktitle}{{\em Proceedings of the 55th Annual Design Automation
  Conference}}. ACM, \bibinfo{pages}{16}.
\newblock


\bibitem[\protect\citeauthoryear{Liu, Wang, Foroosh, Tappen, and Pensky}{Liu
  et~al\mbox{.}}{2015}]%
        {liu2015sparse}
\bibfield{author}{\bibinfo{person}{Baoyuan Liu}, \bibinfo{person}{Min Wang},
  \bibinfo{person}{Hassan Foroosh}, \bibinfo{person}{Marshall Tappen}, {and}
  \bibinfo{person}{Marianna Pensky}.} \bibinfo{year}{2015}\natexlab{}.
\newblock \showarticletitle{Sparse convolutional neural networks}. In
  \bibinfo{booktitle}{{\em Proceedings of the IEEE Conference on Computer
  Vision and Pattern Recognition}}. \bibinfo{pages}{806--814}.
\newblock


\bibitem[\protect\citeauthoryear{Liu, Anguelov, Erhan, Szegedy, Reed, Fu, and
  Berg}{Liu et~al\mbox{.}}{2016a}]%
        {liu2016ssd}
\bibfield{author}{\bibinfo{person}{Wei Liu}, \bibinfo{person}{Dragomir
  Anguelov}, \bibinfo{person}{Dumitru Erhan}, \bibinfo{person}{Christian
  Szegedy}, \bibinfo{person}{Scott Reed}, \bibinfo{person}{Cheng-Yang Fu},
  {and} \bibinfo{person}{Alexander~C Berg}.} \bibinfo{year}{2016}\natexlab{a}.
\newblock \showarticletitle{Ssd: Single shot multibox detector}. In
  \bibinfo{booktitle}{{\em European conference on computer vision}}. Springer,
  \bibinfo{pages}{21--37}.
\newblock


\bibitem[\protect\citeauthoryear{Liu, Dou, Jiang, and Xu}{Liu
  et~al\mbox{.}}{2016b}]%
        {liu2016automatic}
\bibfield{author}{\bibinfo{person}{Zhiqiang Liu}, \bibinfo{person}{Yong Dou},
  \bibinfo{person}{Jingfei Jiang}, {and} \bibinfo{person}{Jinwei Xu}.}
  \bibinfo{year}{2016}\natexlab{b}.
\newblock \showarticletitle{Automatic code generation of convolutional neural
  networks in FPGA implementation}. In \bibinfo{booktitle}{{\em
  Field-Programmable Technology (FPT), 2016 International Conference on}}.
  IEEE, \bibinfo{pages}{61--68}.
\newblock


\bibitem[\protect\citeauthoryear{Lu, Liang, Xiao, and Yan}{Lu
  et~al\mbox{.}}{2017}]%
        {lu2017evaluating}
\bibfield{author}{\bibinfo{person}{Liqiang Lu}, \bibinfo{person}{Yun Liang},
  \bibinfo{person}{Qingcheng Xiao}, {and} \bibinfo{person}{Shengen Yan}.}
  \bibinfo{year}{2017}\natexlab{}.
\newblock \showarticletitle{Evaluating fast algorithms for convolutional neural
  networks on fpgas}. In \bibinfo{booktitle}{{\em Field-Programmable Custom
  Computing Machines (FCCM), 2017 IEEE 25th Annual International Symposium
  on}}. IEEE, \bibinfo{pages}{101--108}.
\newblock


\bibitem[\protect\citeauthoryear{Ma, Cao, Vrudhula, and Seo}{Ma
  et~al\mbox{.}}{2017a}]%
        {ma2017automatic}
\bibfield{author}{\bibinfo{person}{Yufei Ma}, \bibinfo{person}{Yu Cao},
  \bibinfo{person}{Sarma Vrudhula}, {and} \bibinfo{person}{Jae-sun Seo}.}
  \bibinfo{year}{2017}\natexlab{a}.
\newblock \showarticletitle{An automatic RTL compiler for high-throughput FPGA
  implementation of diverse deep convolutional neural networks}. In
  \bibinfo{booktitle}{{\em Field Programmable Logic and Applications (FPL),
  2017 27th International Conference on}}. IEEE, \bibinfo{pages}{1--8}.
\newblock


\bibitem[\protect\citeauthoryear{Ma, Cao, Vrudhula, and Seo}{Ma
  et~al\mbox{.}}{2017b}]%
        {ma2017optimizing}
\bibfield{author}{\bibinfo{person}{Yufei Ma}, \bibinfo{person}{Yu Cao},
  \bibinfo{person}{Sarma Vrudhula}, {and} \bibinfo{person}{Jae-sun Seo}.}
  \bibinfo{year}{2017}\natexlab{b}.
\newblock \showarticletitle{Optimizing Loop Operation and Dataflow in FPGA
  Acceleration of Deep Convolutional Neural Networks}. In
  \bibinfo{booktitle}{{\em Proceedings of the 2017 ACM/SIGDA International
  Symposium on Field-Programmable Gate Arrays}}. ACM, \bibinfo{pages}{45--54}.
\newblock


\bibitem[\protect\citeauthoryear{Mao, Han, Pool, Li, Liu, Wang, and Dally}{Mao
  et~al\mbox{.}}{2017}]%
        {Mao2017Exploring}
\bibfield{author}{\bibinfo{person}{Huizi Mao}, \bibinfo{person}{Song Han},
  \bibinfo{person}{Jeff Pool}, \bibinfo{person}{Wenshuo Li},
  \bibinfo{person}{Xingyu Liu}, \bibinfo{person}{Yu Wang}, {and}
  \bibinfo{person}{William~J. Dally}.} \bibinfo{year}{2017}\natexlab{}.
\newblock \showarticletitle{Exploring the Granularity of Sparsity in
  Convolutional Neural Networks}. In \bibinfo{booktitle}{{\em Computer Vision
  and Pattern Recognition Workshops}}. \bibinfo{pages}{1927--1934}.
\newblock


\bibitem[\protect\citeauthoryear{Morcel, Akkary, Hajj, Saghir, Keshavamurthy,
  Khanna, and Artail}{Morcel et~al\mbox{.}}{2017}]%
        {morcel2017minimalist}
\bibfield{author}{\bibinfo{person}{Raghid Morcel}, \bibinfo{person}{Haitham
  Akkary}, \bibinfo{person}{Hazem Hajj}, \bibinfo{person}{Mazen Saghir},
  \bibinfo{person}{Anil Keshavamurthy}, \bibinfo{person}{Rahul Khanna}, {and}
  \bibinfo{person}{Hassan Artail}.} \bibinfo{year}{2017}\natexlab{}.
\newblock \showarticletitle{Minimalist Design for Accelerating Convolutional
  Neural Networks for Low-End FPGA Platforms}. In \bibinfo{booktitle}{{\em
  Field-Programmable Custom Computing Machines (FCCM), 2017 IEEE 25th Annual
  International Symposium on}}. IEEE, \bibinfo{pages}{196--196}.
\newblock


\bibitem[\protect\citeauthoryear{Moss, Nurvitadhi, Sim, Mishra, Marr,
  Subhaschandra, and Leong}{Moss et~al\mbox{.}}{2017}]%
        {moss2017high}
\bibfield{author}{\bibinfo{person}{Duncan~JM Moss}, \bibinfo{person}{Eriko
  Nurvitadhi}, \bibinfo{person}{Jaewoong Sim}, \bibinfo{person}{Asit Mishra},
  \bibinfo{person}{Debbie Marr}, \bibinfo{person}{Suchit Subhaschandra}, {and}
  \bibinfo{person}{Philip~HW Leong}.} \bibinfo{year}{2017}\natexlab{}.
\newblock \showarticletitle{High performance binary neural networks on the
  Xeon+ FPGA™ platform}. In \bibinfo{booktitle}{{\em Field Programmable Logic
  and Applications (FPL), 2017 27th International Conference on}}. IEEE,
  \bibinfo{pages}{1--4}.
\newblock


\bibitem[\protect\citeauthoryear{Motamedi, Gysel, Akella, and Ghiasi}{Motamedi
  et~al\mbox{.}}{2016}]%
        {motamedi2016design}
\bibfield{author}{\bibinfo{person}{Mohammad Motamedi}, \bibinfo{person}{Philipp
  Gysel}, \bibinfo{person}{Venkatesh Akella}, {and} \bibinfo{person}{Soheil
  Ghiasi}.} \bibinfo{year}{2016}\natexlab{}.
\newblock \showarticletitle{Design space exploration of fpga-based deep
  convolutional neural networks}. In \bibinfo{booktitle}{{\em Design Automation
  Conference (ASP-DAC), 2016 21st Asia and South Pacific}}. IEEE,
  \bibinfo{pages}{575--580}.
\newblock


\bibitem[\protect\citeauthoryear{Nakahara, Fujii, and Sato}{Nakahara
  et~al\mbox{.}}{2017a}]%
        {nakahara2017fully}
\bibfield{author}{\bibinfo{person}{Hiroki Nakahara}, \bibinfo{person}{Tomoya
  Fujii}, {and} \bibinfo{person}{Shimpei Sato}.}
  \bibinfo{year}{2017}\natexlab{a}.
\newblock \showarticletitle{A fully connected layer elimination for a binarizec
  convolutional neural network on an FPGA}. In \bibinfo{booktitle}{{\em Field
  Programmable Logic and Applications (FPL), 2017 27th International Conference
  on}}. IEEE, \bibinfo{pages}{1--4}.
\newblock


\bibitem[\protect\citeauthoryear{Nakahara, Yonekawa, Iwamoto, and
  Motomura}{Nakahara et~al\mbox{.}}{2017b}]%
        {nakahara2017batch}
\bibfield{author}{\bibinfo{person}{Hiroki Nakahara}, \bibinfo{person}{Haruyoshi
  Yonekawa}, \bibinfo{person}{Hisashi Iwamoto}, {and} \bibinfo{person}{Masato
  Motomura}.} \bibinfo{year}{2017}\natexlab{b}.
\newblock \showarticletitle{A Batch Normalization Free Binarized Convolutional
  Deep Neural Network on an FPGA}. In \bibinfo{booktitle}{{\em Proceedings of
  the 2017 ACM/SIGDA International Symposium on Field-Programmable Gate
  Arrays}}. ACM, \bibinfo{pages}{290--290}.
\newblock


\bibitem[\protect\citeauthoryear{Nguyen, Kim, and Lee}{Nguyen
  et~al\mbox{.}}{2017}]%
        {nguyen2017double}
\bibfield{author}{\bibinfo{person}{Dong Nguyen}, \bibinfo{person}{Daewoo Kim},
  {and} \bibinfo{person}{Jongeun Lee}.} \bibinfo{year}{2017}\natexlab{}.
\newblock \showarticletitle{Double MAC: Doubling the performance of
  convolutional neural networks on modern FPGAs}. In \bibinfo{booktitle}{{\em
  2017 Design, Automation \& Test in Europe Conference \& Exhibition (DATE)}}.
  IEEE, \bibinfo{pages}{890--893}.
\newblock


\bibitem[\protect\citeauthoryear{Nurvitadhi, Sheffield, Sim, Mishra, Venkatesh,
  and Marr}{Nurvitadhi et~al\mbox{.}}{2016}]%
        {nurvitadhi2016accelerating}
\bibfield{author}{\bibinfo{person}{Eriko Nurvitadhi}, \bibinfo{person}{David
  Sheffield}, \bibinfo{person}{Jaewoong Sim}, \bibinfo{person}{Asit Mishra},
  \bibinfo{person}{Ganesh Venkatesh}, {and} \bibinfo{person}{Debbie Marr}.}
  \bibinfo{year}{2016}\natexlab{}.
\newblock \showarticletitle{Accelerating Binarized Neural Networks: Comparison
  of FPGA, CPU, GPU, and ASIC}. In \bibinfo{booktitle}{{\em Field-Programmable
  Technology (FPT), 2016 International Conference on}}. IEEE,
  \bibinfo{pages}{77--84}.
\newblock


\bibitem[\protect\citeauthoryear{Podili, Zhang, and Prasanna}{Podili
  et~al\mbox{.}}{2017}]%
        {podili2017fast}
\bibfield{author}{\bibinfo{person}{Abhinav Podili}, \bibinfo{person}{Chi
  Zhang}, {and} \bibinfo{person}{Viktor Prasanna}.}
  \bibinfo{year}{2017}\natexlab{}.
\newblock \showarticletitle{Fast and efficient implementation of Convolutional
  Neural Networks on FPGA}. In \bibinfo{booktitle}{{\em Application-specific
  Systems, Architectures and Processors (ASAP), 2017 IEEE 28th International
  Conference on}}. IEEE, \bibinfo{pages}{11--18}.
\newblock


\bibitem[\protect\citeauthoryear{Prost-Boucle, Bourge, P{\'e}trot, Alemdar,
  Caldwell, and Leroy}{Prost-Boucle et~al\mbox{.}}{2017}]%
        {prost2017scalable}
\bibfield{author}{\bibinfo{person}{Adrien Prost-Boucle}, \bibinfo{person}{Alban
  Bourge}, \bibinfo{person}{Fr{\'e}d{\'e}ric P{\'e}trot},
  \bibinfo{person}{Hande Alemdar}, \bibinfo{person}{Nicholas Caldwell}, {and}
  \bibinfo{person}{Vincent Leroy}.} \bibinfo{year}{2017}\natexlab{}.
\newblock \showarticletitle{Scalable high-performance architecture for
  convolutional ternary neural networks on FPGA}. In \bibinfo{booktitle}{{\em
  Field Programmable Logic and Applications (FPL), 2017 27th International
  Conference on}}. IEEE, \bibinfo{pages}{1--7}.
\newblock


\bibitem[\protect\citeauthoryear{Qiu, Wang, Yao, Guo, Li, Zhou, Yu, Tang, Xu,
  Song, et~al\mbox{.}}{Qiu et~al\mbox{.}}{2016}]%
        {qiu2016going}
\bibfield{author}{\bibinfo{person}{Jiantao Qiu}, \bibinfo{person}{Jie Wang},
  \bibinfo{person}{Song Yao}, \bibinfo{person}{Kaiyuan Guo},
  \bibinfo{person}{Boxun Li}, \bibinfo{person}{Erjin Zhou},
  \bibinfo{person}{Jincheng Yu}, \bibinfo{person}{Tianqi Tang},
  \bibinfo{person}{Ningyi Xu}, \bibinfo{person}{Sen Song}, {et~al\mbox{.}}}
  \bibinfo{year}{2016}\natexlab{}.
\newblock \showarticletitle{Going deeper with embedded fpga platform for
  convolutional neural network}. In \bibinfo{booktitle}{{\em Proceedings of the
  2016 ACM/SIGDA International Symposium on Field-Programmable Gate Arrays}}.
  ACM, \bibinfo{pages}{26--35}.
\newblock


\bibitem[\protect\citeauthoryear{Russakovsky, Deng, Su, Krause, Satheesh, Ma,
  Huang, Karpathy, Khosla, Bernstein, Berg, and Fei-Fei}{Russakovsky
  et~al\mbox{.}}{2015}]%
        {ILSVRC15}
\bibfield{author}{\bibinfo{person}{Olga Russakovsky}, \bibinfo{person}{Jia
  Deng}, \bibinfo{person}{Hao Su}, \bibinfo{person}{Jonathan Krause},
  \bibinfo{person}{Sanjeev Satheesh}, \bibinfo{person}{Sean Ma},
  \bibinfo{person}{Zhiheng Huang}, \bibinfo{person}{Andrej Karpathy},
  \bibinfo{person}{Aditya Khosla}, \bibinfo{person}{Michael Bernstein},
  \bibinfo{person}{Alexander~C. Berg}, {and} \bibinfo{person}{Li Fei-Fei}.}
  \bibinfo{year}{2015}\natexlab{}.
\newblock \showarticletitle{{ImageNet Large Scale Visual Recognition
  Challenge}}.
\newblock \bibinfo{journal}{{\em International Journal of Computer Vision
  (IJCV)\/}} \bibinfo{volume}{115}, \bibinfo{number}{3} (\bibinfo{year}{2015}),
  \bibinfo{pages}{211--252}.
\newblock
\showDOI{%
\url{https://doi.org/10.1007/s11263-015-0816-y}}


\bibitem[\protect\citeauthoryear{Samragh, Ghasemzadeh, and Koushanfar}{Samragh
  et~al\mbox{.}}{2017}]%
        {samragh2017customizing}
\bibfield{author}{\bibinfo{person}{Mohammad Samragh}, \bibinfo{person}{Mohammad
  Ghasemzadeh}, {and} \bibinfo{person}{Farinaz Koushanfar}.}
  \bibinfo{year}{2017}\natexlab{}.
\newblock \showarticletitle{Customizing neural networks for efficient fpga
  implementation}. In \bibinfo{booktitle}{{\em Field-Programmable Custom
  Computing Machines (FCCM), 2017 IEEE 25th Annual International Symposium
  on}}. IEEE, \bibinfo{pages}{85--92}.
\newblock


\bibitem[\protect\citeauthoryear{Sharma, Park, Mahajan, Amaro, Kim, Shao,
  Mishra, and Esmaeilzadeh}{Sharma et~al\mbox{.}}{2016}]%
        {sharma2016high}
\bibfield{author}{\bibinfo{person}{Hardik Sharma}, \bibinfo{person}{Jongse
  Park}, \bibinfo{person}{Divya Mahajan}, \bibinfo{person}{Emmanuel Amaro},
  \bibinfo{person}{Joon~Kyung Kim}, \bibinfo{person}{Chenkai Shao},
  \bibinfo{person}{Asit Mishra}, {and} \bibinfo{person}{Hadi Esmaeilzadeh}.}
  \bibinfo{year}{2016}\natexlab{}.
\newblock \showarticletitle{From high-level deep neural models to FPGAs}. In
  \bibinfo{booktitle}{{\em Microarchitecture (MICRO), 2016 49th Annual IEEE/ACM
  International Symposium on}}. IEEE, \bibinfo{pages}{1--12}.
\newblock


\bibitem[\protect\citeauthoryear{Shen, Huang, Wang, Qiao, Wen, and Zhang}{Shen
  et~al\mbox{.}}{2018}]%
        {Shen2018Towards}
\bibfield{author}{\bibinfo{person}{Junzhong Shen}, \bibinfo{person}{You Huang},
  \bibinfo{person}{Zelong Wang}, \bibinfo{person}{Yuran Qiao},
  \bibinfo{person}{Mei Wen}, {and} \bibinfo{person}{Chunyuan Zhang}.}
  \bibinfo{year}{2018}\natexlab{}.
\newblock \showarticletitle{Towards a Uniform Template-based Architecture for
  Accelerating 2D and 3D CNNs on FPGA}. In \bibinfo{booktitle}{{\em Acm/sigda
  International Symposium}}. \bibinfo{pages}{97--106}.
\newblock


\bibitem[\protect\citeauthoryear{Shen, Ferdman, and Milder}{Shen
  et~al\mbox{.}}{2016}]%
        {shen2016overcoming}
\bibfield{author}{\bibinfo{person}{Yongming Shen}, \bibinfo{person}{Michael
  Ferdman}, {and} \bibinfo{person}{Peter Milder}.}
  \bibinfo{year}{2016}\natexlab{}.
\newblock \showarticletitle{Overcoming resource underutilization in spatial CNN
  accelerators}. In \bibinfo{booktitle}{{\em Field Programmable Logic and
  Applications (FPL), 2016 26th International Conference on}}. IEEE,
  \bibinfo{pages}{1--4}.
\newblock


\bibitem[\protect\citeauthoryear{Shen, Ferdman, and Milder}{Shen
  et~al\mbox{.}}{2017}]%
        {shen2017escher}
\bibfield{author}{\bibinfo{person}{Yongming Shen}, \bibinfo{person}{Michael
  Ferdman}, {and} \bibinfo{person}{Peter Milder}.}
  \bibinfo{year}{2017}\natexlab{}.
\newblock \showarticletitle{Escher: A CNN Accelerator with Flexible Buffering
  to Minimize Off-Chip Transfer}. In \bibinfo{booktitle}{{\em Proceedings of
  the 25th IEEE International Symposium on Field-Programmable Custom Computing
  Machines (FCCM’17). IEEE Computer Society, Los Alamitos, CA, USA}}.
\newblock


\bibitem[\protect\citeauthoryear{Simonyan and Zisserman}{Simonyan and
  Zisserman}{2014}]%
        {simonyan2014very}
\bibfield{author}{\bibinfo{person}{Karen Simonyan} {and}
  \bibinfo{person}{Andrew Zisserman}.} \bibinfo{year}{2014}\natexlab{}.
\newblock \showarticletitle{Very deep convolutional networks for large-scale
  image recognition}.
\newblock \bibinfo{journal}{{\em arXiv preprint arXiv:1409.1556\/}}
  (\bibinfo{year}{2014}).
\newblock


\bibitem[\protect\citeauthoryear{Suda, Chandra, Dasika, Mohanty, Ma, Vrudhula,
  Seo, and Cao}{Suda et~al\mbox{.}}{2016}]%
        {suda2016throughput}
\bibfield{author}{\bibinfo{person}{Naveen Suda}, \bibinfo{person}{Vikas
  Chandra}, \bibinfo{person}{Ganesh Dasika}, \bibinfo{person}{Abinash Mohanty},
  \bibinfo{person}{Yufei Ma}, \bibinfo{person}{Sarma Vrudhula},
  \bibinfo{person}{Jae-sun Seo}, {and} \bibinfo{person}{Yu Cao}.}
  \bibinfo{year}{2016}\natexlab{}.
\newblock \showarticletitle{Throughput-optimized OpenCL-based FPGA accelerator
  for large-scale convolutional neural networks}. In \bibinfo{booktitle}{{\em
  Proceedings of the 2016 ACM/SIGDA International Symposium on
  Field-Programmable Gate Arrays}}. ACM, \bibinfo{pages}{16--25}.
\newblock


\bibitem[\protect\citeauthoryear{Szegedy, Liu, Jia, Sermanet, Reed, Anguelov,
  Erhan, Vanhoucke, Rabinovich, et~al\mbox{.}}{Szegedy et~al\mbox{.}}{2015}]%
        {szegedy2015going}
\bibfield{author}{\bibinfo{person}{Christian Szegedy}, \bibinfo{person}{Wei
  Liu}, \bibinfo{person}{Yangqing Jia}, \bibinfo{person}{Pierre Sermanet},
  \bibinfo{person}{Scott Reed}, \bibinfo{person}{Dragomir Anguelov},
  \bibinfo{person}{Dumitru Erhan}, \bibinfo{person}{Vincent Vanhoucke},
  \bibinfo{person}{Andrew Rabinovich}, {et~al\mbox{.}}}
  \bibinfo{year}{2015}\natexlab{}.
\newblock \showarticletitle{Going deeper with convolutions}. Cvpr.
\newblock


\bibitem[\protect\citeauthoryear{Tan, Chen, Pang, Vasudevan, and Le}{Tan
  et~al\mbox{.}}{2018}]%
        {tan2018mnasnet}
\bibfield{author}{\bibinfo{person}{Mingxing Tan}, \bibinfo{person}{Bo Chen},
  \bibinfo{person}{Ruoming Pang}, \bibinfo{person}{Vijay Vasudevan}, {and}
  \bibinfo{person}{Quoc~V Le}.} \bibinfo{year}{2018}\natexlab{}.
\newblock \showarticletitle{Mnasnet: Platform-aware neural architecture search
  for mobile}.
\newblock \bibinfo{journal}{{\em arXiv preprint arXiv:1807.11626\/}}
  (\bibinfo{year}{2018}).
\newblock


\bibitem[\protect\citeauthoryear{Umuroglu, Fraser, Gambardella, Blott, Leong,
  Jahre, and Vissers}{Umuroglu et~al\mbox{.}}{2017}]%
        {umuroglu2017finn}
\bibfield{author}{\bibinfo{person}{Yaman Umuroglu}, \bibinfo{person}{Nicholas~J
  Fraser}, \bibinfo{person}{Giulio Gambardella}, \bibinfo{person}{Michaela
  Blott}, \bibinfo{person}{Philip Leong}, \bibinfo{person}{Magnus Jahre}, {and}
  \bibinfo{person}{Kees Vissers}.} \bibinfo{year}{2017}\natexlab{}.
\newblock \showarticletitle{Finn: A framework for fast, scalable binarized
  neural network inference}. In \bibinfo{booktitle}{{\em Proceedings of the
  2017 ACM/SIGDA International Symposium on Field-Programmable Gate Arrays}}.
  ACM, \bibinfo{pages}{65--74}.
\newblock


\bibitem[\protect\citeauthoryear{Venieris and Bouganis}{Venieris and
  Bouganis}{2017a}]%
        {venieris2017fpgaconvnet}
\bibfield{author}{\bibinfo{person}{Stylianos~I Venieris} {and}
  \bibinfo{person}{Christos-Savvas Bouganis}.}
  \bibinfo{year}{2017}\natexlab{a}.
\newblock \showarticletitle{fpgaConvNet: Automated Mapping of Convolutional
  Neural Networks on FPGAs}. In \bibinfo{booktitle}{{\em Proceedings of the
  2017 ACM/SIGDA International Symposium on Field-Programmable Gate Arrays}}.
  ACM, \bibinfo{pages}{291--292}.
\newblock


\bibitem[\protect\citeauthoryear{Venieris and Bouganis}{Venieris and
  Bouganis}{2017b}]%
        {venieris2017latency}
\bibfield{author}{\bibinfo{person}{Stylianos~I Venieris} {and}
  \bibinfo{person}{Christos-Savvas Bouganis}.}
  \bibinfo{year}{2017}\natexlab{b}.
\newblock \showarticletitle{Latency-driven design for FPGA-based convolutional
  neural networks}. In \bibinfo{booktitle}{{\em Field Programmable Logic and
  Applications (FPL), 2017 27th International Conference on}}. IEEE,
  \bibinfo{pages}{1--8}.
\newblock


\bibitem[\protect\citeauthoryear{Venieris, Kouris, and Bouganis}{Venieris
  et~al\mbox{.}}{2018}]%
        {venieris2018toolflows}
\bibfield{author}{\bibinfo{person}{Stylianos~I Venieris},
  \bibinfo{person}{Alexandros Kouris}, {and} \bibinfo{person}{Christos-Savvas
  Bouganis}.} \bibinfo{year}{2018}\natexlab{}.
\newblock \showarticletitle{Toolflows for Mapping Convolutional Neural Networks
  on FPGAs: A Survey and Future Directions}.
\newblock \bibinfo{journal}{{\em ACM Computing Surveys (CSUR)\/}}
  \bibinfo{volume}{51}, \bibinfo{number}{3} (\bibinfo{year}{2018}),
  \bibinfo{pages}{56}.
\newblock


\bibitem[\protect\citeauthoryear{Wang, Lou, Zhang, Zhu, Lin, and Chen}{Wang
  et~al\mbox{.}}{2018}]%
        {wang2018design}
\bibfield{author}{\bibinfo{person}{Junsong Wang}, \bibinfo{person}{Qiuwen Lou},
  \bibinfo{person}{Xiaofan Zhang}, \bibinfo{person}{Chao Zhu},
  \bibinfo{person}{Yonghua Lin}, {and} \bibinfo{person}{Deming Chen}.}
  \bibinfo{year}{2018}\natexlab{}.
\newblock \showarticletitle{Design Flow of Accelerating Hybrid Extremely Low
  Bit-width Neural Network in Embedded FPGA}.
\newblock \bibinfo{journal}{{\em arXiv preprint arXiv:1808.04311\/}}
  (\bibinfo{year}{2018}).
\newblock


\bibitem[\protect\citeauthoryear{Wang, Yu, Dou, and Gonzalez}{Wang
  et~al\mbox{.}}{2017}]%
        {wang2017skipnet}
\bibfield{author}{\bibinfo{person}{Xin Wang}, \bibinfo{person}{Fisher Yu},
  \bibinfo{person}{Zi-Yi Dou}, {and} \bibinfo{person}{Joseph~E Gonzalez}.}
  \bibinfo{year}{2017}\natexlab{}.
\newblock \showarticletitle{Skipnet: Learning dynamic routing in convolutional
  networks}.
\newblock \bibinfo{journal}{{\em arXiv preprint arXiv:1711.09485\/}}
  (\bibinfo{year}{2017}).
\newblock


\bibitem[\protect\citeauthoryear{Wang, Xu, Han, Li, and Li}{Wang
  et~al\mbox{.}}{2016}]%
        {wang2016deepburning}
\bibfield{author}{\bibinfo{person}{Ying Wang}, \bibinfo{person}{Jie Xu},
  \bibinfo{person}{Yinhe Han}, \bibinfo{person}{Huawei Li}, {and}
  \bibinfo{person}{Xiaowei Li}.} \bibinfo{year}{2016}\natexlab{}.
\newblock \showarticletitle{DeepBurning: automatic generation of FPGA-based
  learning accelerators for the neural network family}. In
  \bibinfo{booktitle}{{\em Design Automation Conference (DAC), 2016 53nd
  ACM/EDAC/IEEE}}. IEEE, \bibinfo{pages}{1--6}.
\newblock


\bibitem[\protect\citeauthoryear{Wei, Yu, Zhang, Chen, Wang, Hu, Liang, and
  Cong}{Wei et~al\mbox{.}}{2017}]%
        {wei2017automated}
\bibfield{author}{\bibinfo{person}{Xuechao Wei}, \bibinfo{person}{Cody~Hao Yu},
  \bibinfo{person}{Peng Zhang}, \bibinfo{person}{Youxiang Chen},
  \bibinfo{person}{Yuxin Wang}, \bibinfo{person}{Han Hu}, \bibinfo{person}{Yun
  Liang}, {and} \bibinfo{person}{Jason Cong}.} \bibinfo{year}{2017}\natexlab{}.
\newblock \showarticletitle{Automated Systolic Array Architecture Synthesis for
  High Throughput CNN Inference on FPGAs}. In \bibinfo{booktitle}{{\em
  Proceedings of the 54th Annual Design Automation Conference 2017}}. ACM,
  \bibinfo{pages}{29}.
\newblock


\bibitem[\protect\citeauthoryear{Winograd}{Winograd}{1980}]%
        {winograd1980arithmetic}
\bibfield{author}{\bibinfo{person}{Shmuel Winograd}.}
  \bibinfo{year}{1980}\natexlab{}.
\newblock \bibinfo{booktitle}{{\em Arithmetic complexity of computations}}.
  Vol.~\bibinfo{volume}{33}.
\newblock \bibinfo{publisher}{Siam}.
\newblock


\bibitem[\protect\citeauthoryear{Wu, Zhang, Berman, and Cho}{Wu
  et~al\mbox{.}}{2017}]%
        {wu2017high}
\bibfield{author}{\bibinfo{person}{Ephrem Wu}, \bibinfo{person}{Xiaoqian
  Zhang}, \bibinfo{person}{David Berman}, {and} \bibinfo{person}{Inkeun Cho}.}
  \bibinfo{year}{2017}\natexlab{}.
\newblock \showarticletitle{A high-throughput reconfigurable processing array
  for neural networks}. In \bibinfo{booktitle}{{\em Field Programmable Logic
  and Applications (FPL), 2017 27th International Conference on}}. IEEE,
  \bibinfo{pages}{1--4}.
\newblock


\bibitem[\protect\citeauthoryear{Xiao, Liang, Lu, Yan, and Tai}{Xiao
  et~al\mbox{.}}{2017}]%
        {xiao2017exploring}
\bibfield{author}{\bibinfo{person}{Qingcheng Xiao}, \bibinfo{person}{Yun
  Liang}, \bibinfo{person}{Liqiang Lu}, \bibinfo{person}{Shengen Yan}, {and}
  \bibinfo{person}{Yu-Wing Tai}.} \bibinfo{year}{2017}\natexlab{}.
\newblock \showarticletitle{Exploring Heterogeneous Algorithms for Accelerating
  Deep Convolutional Neural Networks on FPGAs}. In \bibinfo{booktitle}{{\em
  Proceedings of the 54th Annual Design Automation Conference 2017}}. ACM,
  \bibinfo{pages}{62}.
\newblock


\bibitem[\protect\citeauthoryear{Yang, He, and Fan}{Yang et~al\mbox{.}}{2018}]%
        {yang2018fully}
\bibfield{author}{\bibinfo{person}{Li Yang}, \bibinfo{person}{Zhezhi He}, {and}
  \bibinfo{person}{Deliang Fan}.} \bibinfo{year}{2018}\natexlab{}.
\newblock \showarticletitle{A Fully Onchip Binarized Convolutional Neural
  Network FPGA Impelmentation with Accurate Inference}. In
  \bibinfo{booktitle}{{\em Proceedings of the International Symposium on Low
  Power Electronics and Design}}. ACM, \bibinfo{pages}{50}.
\newblock


\bibitem[\protect\citeauthoryear{Yu, Hu, Ning, Qiu, Guo, Wang, and Yang}{Yu
  et~al\mbox{.}}{2017}]%
        {Yu2017Instruction}
\bibfield{author}{\bibinfo{person}{Jincheng Yu}, \bibinfo{person}{Yiming Hu},
  \bibinfo{person}{Xuefei Ning}, \bibinfo{person}{Jiantao Qiu},
  \bibinfo{person}{Kaiyuan Guo}, \bibinfo{person}{Yu Wang}, {and}
  \bibinfo{person}{Huazhong Yang}.} \bibinfo{year}{2017}\natexlab{}.
\newblock \showarticletitle{Instruction driven cross-layer CNN accelerator with
  winograd transformation on FPGA}. In \bibinfo{booktitle}{{\em International
  Conference on Field Programmable Technology}}. \bibinfo{pages}{227--230}.
\newblock


\bibitem[\protect\citeauthoryear{Zhang, Fang, Zhou, Pan, and Cong}{Zhang
  et~al\mbox{.}}{2016a}]%
        {zhang2016caffeine}
\bibfield{author}{\bibinfo{person}{Chen Zhang}, \bibinfo{person}{Zhenman Fang},
  \bibinfo{person}{Peipei Zhou}, \bibinfo{person}{Peichen Pan}, {and}
  \bibinfo{person}{Jason Cong}.} \bibinfo{year}{2016}\natexlab{a}.
\newblock \showarticletitle{Caffeine: Towards uniformed representation and
  acceleration for deep convolutional neural networks}. In
  \bibinfo{booktitle}{{\em Computer-Aided Design (ICCAD), 2016 IEEE/ACM
  International Conference on}}. IEEE, \bibinfo{pages}{1--8}.
\newblock


\bibitem[\protect\citeauthoryear{Zhang, Li, Sun, Guan, Xiao, and Cong}{Zhang
  et~al\mbox{.}}{2015a}]%
        {zhang2015optimizing}
\bibfield{author}{\bibinfo{person}{Chen Zhang}, \bibinfo{person}{Peng Li},
  \bibinfo{person}{Guangyu Sun}, \bibinfo{person}{Yijin Guan},
  \bibinfo{person}{Bingjun Xiao}, {and} \bibinfo{person}{Jason Cong}.}
  \bibinfo{year}{2015}\natexlab{a}.
\newblock \showarticletitle{Optimizing fpga-based accelerator design for deep
  convolutional neural networks}. In \bibinfo{booktitle}{{\em Proceedings of
  the 2015 ACM/SIGDA International Symposium on Field-Programmable Gate
  Arrays}}. ACM, \bibinfo{pages}{161--170}.
\newblock


\bibitem[\protect\citeauthoryear{Zhang and Prasanna}{Zhang and
  Prasanna}{2017}]%
        {zhang2017frequency}
\bibfield{author}{\bibinfo{person}{Chi Zhang} {and} \bibinfo{person}{Viktor
  Prasanna}.} \bibinfo{year}{2017}\natexlab{}.
\newblock \showarticletitle{Frequency domain acceleration of convolutional
  neural networks on CPU-FPGA shared memory system}. In
  \bibinfo{booktitle}{{\em Proceedings of the 2017 ACM/SIGDA International
  Symposium on Field-Programmable Gate Arrays}}. ACM, \bibinfo{pages}{35--44}.
\newblock


\bibitem[\protect\citeauthoryear{Zhang, Wu, Sun, Sun, Luo, and Cong}{Zhang
  et~al\mbox{.}}{2016b}]%
        {zhang2016energy}
\bibfield{author}{\bibinfo{person}{Chen Zhang}, \bibinfo{person}{Di Wu},
  \bibinfo{person}{Jiayu Sun}, \bibinfo{person}{Guangyu Sun},
  \bibinfo{person}{Guojie Luo}, {and} \bibinfo{person}{Jason Cong}.}
  \bibinfo{year}{2016}\natexlab{b}.
\newblock \showarticletitle{Energy-Efficient CNN Implementation on a Deeply
  Pipelined FPGA Cluster}. In \bibinfo{booktitle}{{\em Proceedings of the 2016
  International Symposium on Low Power Electronics and Design}}. ACM,
  \bibinfo{pages}{326--331}.
\newblock


\bibitem[\protect\citeauthoryear{Zhang and Li}{Zhang and Li}{2017}]%
        {zhang2017improving}
\bibfield{author}{\bibinfo{person}{Jialiang Zhang} {and} \bibinfo{person}{Jing
  Li}.} \bibinfo{year}{2017}\natexlab{}.
\newblock \showarticletitle{Improving the Performance of OpenCL-based FPGA
  Accelerator for Convolutional Neural Network.}. In \bibinfo{booktitle}{{\em
  FPGA}}. \bibinfo{pages}{25--34}.
\newblock


\bibitem[\protect\citeauthoryear{Zhang, Wang, Zhu, Lin, Xiong, Hwu, and
  Chen}{Zhang et~al\mbox{.}}{2018}]%
        {zhang2018dnnbuilder}
\bibfield{author}{\bibinfo{person}{Xiaofan Zhang}, \bibinfo{person}{Junsong
  Wang}, \bibinfo{person}{Chao Zhu}, \bibinfo{person}{Yonghua Lin},
  \bibinfo{person}{Jinjun Xiong}, \bibinfo{person}{Wen-mei Hwu}, {and}
  \bibinfo{person}{Deming Chen}.} \bibinfo{year}{2018}\natexlab{}.
\newblock \showarticletitle{DNNBuilder: an automated tool for building
  high-performance DNN hardware accelerators for FPGAs}. In
  \bibinfo{booktitle}{{\em Proceedings of the International Conference on
  Computer-Aided Design}}. ACM, \bibinfo{pages}{56}.
\newblock


\bibitem[\protect\citeauthoryear{Zhang, Zhou, Lin, and Sun}{Zhang
  et~al\mbox{.}}{2017}]%
        {zhang2017shufflenet}
\bibfield{author}{\bibinfo{person}{Xiangyu Zhang}, \bibinfo{person}{Xinyu
  Zhou}, \bibinfo{person}{Mengxiao Lin}, {and} \bibinfo{person}{Jian Sun}.}
  \bibinfo{year}{2017}\natexlab{}.
\newblock \showarticletitle{ShuffleNet: An Extremely Efficient Convolutional
  Neural Network for Mobile Devices}.
\newblock \bibinfo{journal}{{\em CoRR\/}}  \bibinfo{volume}{abs/1707.01083}
  (\bibinfo{year}{2017}).
\newblock
\showeprint[arxiv]{1707.01083}
\showURL{%
\url{http://arxiv.org/abs/1707.01083}}


\bibitem[\protect\citeauthoryear{Zhang, Zou, Ming, He, and Sun}{Zhang
  et~al\mbox{.}}{2015b}]%
        {zhang2015efficient}
\bibfield{author}{\bibinfo{person}{Xiangyu Zhang}, \bibinfo{person}{Jianhua
  Zou}, \bibinfo{person}{Xiang Ming}, \bibinfo{person}{Kaiming He}, {and}
  \bibinfo{person}{Jian Sun}.} \bibinfo{year}{2015}\natexlab{b}.
\newblock \showarticletitle{Efficient and accurate approximations of nonlinear
  convolutional networks}. In \bibinfo{booktitle}{{\em Proceedings of the IEEE
  Conference on Computer Vision and Pattern Recognition}}.
  \bibinfo{pages}{1984--1992}.
\newblock


\bibitem[\protect\citeauthoryear{Zhao, Song, Zhang, Xing, Lin, Srivastava,
  Gupta, and Zhang}{Zhao et~al\mbox{.}}{2017}]%
        {zhao2017accelerating}
\bibfield{author}{\bibinfo{person}{Ritchie Zhao}, \bibinfo{person}{Weinan
  Song}, \bibinfo{person}{Wentao Zhang}, \bibinfo{person}{Tianwei Xing},
  \bibinfo{person}{Jeng-Hau Lin}, \bibinfo{person}{Mani~B Srivastava},
  \bibinfo{person}{Rajesh Gupta}, {and} \bibinfo{person}{Zhiru Zhang}.}
  \bibinfo{year}{2017}\natexlab{}.
\newblock \showarticletitle{Accelerating Binarized Convolutional Neural
  Networks with Software-Programmable FPGAs.}. In \bibinfo{booktitle}{{\em
  FPGA}}. \bibinfo{pages}{15--24}.
\newblock


\bibitem[\protect\citeauthoryear{Zhou, Wu, Ni, Zhou, Wen, and Zou}{Zhou
  et~al\mbox{.}}{2016}]%
        {zhou2016dorefa}
\bibfield{author}{\bibinfo{person}{Shuchang Zhou}, \bibinfo{person}{Yuxin Wu},
  \bibinfo{person}{Zekun Ni}, \bibinfo{person}{Xinyu Zhou}, \bibinfo{person}{He
  Wen}, {and} \bibinfo{person}{Yuheng Zou}.} \bibinfo{year}{2016}\natexlab{}.
\newblock \showarticletitle{DoReFa-Net: Training low bitwidth convolutional
  neural networks with low bitwidth gradients}.
\newblock \bibinfo{journal}{{\em arXiv preprint arXiv:1606.06160\/}}
  (\bibinfo{year}{2016}).
\newblock


\bibitem[\protect\citeauthoryear{Zhu, Han, Mao, and Dally}{Zhu
  et~al\mbox{.}}{2016}]%
        {zhu2016trained}
\bibfield{author}{\bibinfo{person}{Chenzhuo Zhu}, \bibinfo{person}{Song Han},
  \bibinfo{person}{Huizi Mao}, {and} \bibinfo{person}{William~J Dally}.}
  \bibinfo{year}{2016}\natexlab{}.
\newblock \showarticletitle{Trained ternary quantization}.
\newblock \bibinfo{journal}{{\em arXiv preprint arXiv:1612.01064\/}}
  (\bibinfo{year}{2016}).
\newblock


\bibitem[\protect\citeauthoryear{Zhuge, Liu, Zhang, Gummadi, Xiong, and
  Chen}{Zhuge et~al\mbox{.}}{2018}]%
        {zhuge2018face}
\bibfield{author}{\bibinfo{person}{Chuanhao Zhuge}, \bibinfo{person}{Xinheng
  Liu}, \bibinfo{person}{Xiaofan Zhang}, \bibinfo{person}{Sudeep Gummadi},
  \bibinfo{person}{Jinjun Xiong}, {and} \bibinfo{person}{Deming Chen}.}
  \bibinfo{year}{2018}\natexlab{}.
\newblock \showarticletitle{Face Recognition with Hybrid Efficient Convolution
  Algorithms on FPGAs}. In \bibinfo{booktitle}{{\em Proceedings of the 2018 on
  Great Lakes Symposium on VLSI}}. ACM, \bibinfo{pages}{123--128}.
\newblock


\end{thebibliography}

\end{document}